\documentclass[conference]{IEEEtran}
\IEEEoverridecommandlockouts
\usepackage{cite}
\usepackage{amsmath,amssymb,amsfonts}
\usepackage{algorithmic}
\usepackage{graphicx}
\usepackage{textcomp}
\usepackage{xcolor}

\usepackage{amsthm}
\usepackage{fontawesome}
\usepackage{setspace}

\usepackage{amsthm}
\usepackage{pifont}
\usepackage{enumitem}

\usepackage{subcaption}

\usepackage{caption}
\usepackage{multirow}
\usepackage{booktabs}
\usepackage{array}
\usepackage{bm}
\usepackage{adjustbox}

\theoremstyle{plain}
\newtheorem{theorem}{Theorem}
\usepackage{enumitem}
\usepackage{lipsum} % for generating dummy text

\theoremstyle{plain}
\newtheorem{definition}[theorem]{Definition}
\theoremstyle{plain}
\newtheorem{lemma}[theorem]{Lemma}

\theoremstyle{plain}
\newtheorem{example}[theorem]{Example} % 定义 example 环境，编号按 section

\theoremstyle{plain}
\newenvironment{prf}[1][Proof]{\noindent{#1.} }{\par\leavevmode}

\def\BibTeX{{\rm B\kern-.05em{\sc i\kern-.025em b}\kern-.08em
    T\kern-.1667em\lower.7ex\hbox{E}\kern-.125emX}}

\def\tech{0} % 1是technical report

\begin{document}

% \title{\parbox{1.2\linewidth}{\centering LH-plugin: Overcoming Triangle Inequality Constraints in
% Trajectory Representation Learning for Similarity Computation}}
% \title{Effective Trajectory Representation Learning: Triangle Inequality Violations in Similarity Computation}

% \title{LH-plugin: Overcoming Triangle Inequality Constraints of Distance Metrics for
% Trajectory Representation Learning in Similarity Computation}

\title{\vspace{-0.05in}Towards Robust Trajectory Embedding for Similarity Computation: When Triangle Inequality Violations in Distance Metrics Matter \thanks{This paper was supported by NSF of China(62032003, 62425203, 62372061), Shenzhen Science and Technology Program JCYJ 20240813165400002, Young Elite Scientists Sponsorship Program by CAST under Grant 2023QNRC001 and State Key Laboratory Of Networking And Switching Technology. \faEnvelopeO{} Corresponding author.}\vspace{-0.05in}
}

% \title{LH-plugin: Liberating Trajectory Representation Learning from Triangle Inequality Constraints in Similarity Computation}

% \title{Trajectory Representation Learning for Similarity Computation Revisited: When Triangle Inequality Violation Matter}

% \author{
% \IEEEauthorblockN{Jianing Si}
% \IEEEauthorblockA{\textit{BUPT, China} \\
% sijianing@bupt.edu.cn}
% \and
% \IEEEauthorblockN{Haitao Yuan}
% \IEEEauthorblockA{\textit{NTU, Singapore} \\
% haitao.yuan@ntu.edu.sg}
% \and
% \IEEEauthorblockN{Nan Jiang}
% \IEEEauthorblockA{\textit{BUPT, China} \\
% jn_bupt@bupt.edu.cn}

% \linebreakand
% \IEEEauthorblockN{Minxiao Chen}
% \IEEEauthorblockA{\textit{BUPT, China} \\
% chenminxiao@bupt.edu.cn}
% \and
% \IEEEauthorblockN{Xiao Ma}
% \IEEEauthorblockA{\textit{BUPT, China} \\
% maxiao18@bupt.edu.cn}
% \and
% \IEEEauthorblockN{Shangguang Wang}
% \IEEEauthorblockA{\textit{BUPT, China} \\
% sgwang@bupt.edu.cn}
% \and
% }

\author{
\IEEEauthorblockN{Jianing Si\IEEEauthorrefmark{1}\IEEEauthorrefmark{3}, Haitao Yuan\faEnvelopeO{}\IEEEauthorrefmark{2},  Nan Jiang\IEEEauthorrefmark{1}, Minxiao Chen\IEEEauthorrefmark{1}, Xiao Ma\IEEEauthorrefmark{1}, Shangguang Wang\IEEEauthorrefmark{1}}
\IEEEauthorblockA{\IEEEauthorrefmark{1}\textit{BUPT, China}\qquad \IEEEauthorrefmark{2}\textit{NTU, Singapore} \qquad \IEEEauthorrefmark{3}Beiyou Shenzhen Institute, China\\
\vspace{-0.02in}
\IEEEauthorrefmark{1}\{sijianing, jn\_bupt, chenminxiao, maxiao18, sgwang\}@bupt.edu.cn, \qquad \IEEEauthorrefmark{2}haitao.yuan@ntu.edu.sg}
\vspace{-0.2in}
}

\maketitle

\begin{abstract}
Trajectory similarity is a cornerstone of trajectory data management and analysis. Traditional similarity functions often suffer from high computational complexity and a reliance on specific distance metrics, prompting a shift towards deep representation learning in Euclidean space. However, existing Euclidean-based trajectory embeddings often face challenges due to the triangle inequality constraints that do not universally hold for trajectory data. To address this issue, this paper introduces a novel approach by incorporating non-Euclidean geometry, specifically hyperbolic space, into trajectory representation learning. We present the first-ever integration of hyperbolic space to resolve the inherent limitations of the triangle inequality in Euclidean embeddings. In particular, we achieve it by designing a Lorentz distance measure, which is proven to overcome triangle inequality constraints.
Additionally, we design a model-agnostic framework \textbf{LH}-plugin to seamlessly integrate hyperbolic embeddings into existing representation learning pipelines. This includes a novel projection method optimized with the Cosh function to prevent the diminishment of distances, supported by a theoretical foundation. Furthermore, we propose a dynamic fusion distance that intelligently adapts to variations in triangle inequality constraints across different trajectory pairs, blending Lorentzian and Euclidean distances for more robust similarity calculations. Comprehensive experimental evaluations demonstrate that our approach effectively enhances the accuracy of trajectory similarity measures in state-of-the-art models across multiple real-world datasets. The \textbf{LH}-plugin not only addresses the triangle inequality issues but also significantly refines the precision of trajectory similarity computations, marking a substantial advancement in the field of trajectory representation learning.
\end{abstract}

\vspace{-0.03in}
\section{Introduction}
Computing trajectory similarity distance is crucial for various tasks in trajectory data management and analytics, as referenced in several studies~\cite{survey1,survey2,survey3,survey4,survey5, yuan2, yuan11}. In particular, trajectory similarity metrics effectively measure the distance between two distinct trajectories, which is vital for many downstream applications such as trajectory clustering~\cite{traj_cluster, traj_cluster2, traj_cluster3, cluster_survey}, efficient trajectory querying and retrieval~\cite{sim_query, retrieval2, yuan4, yuan7, yuan10}, and the optimization of traffic conditions~\cite{traj_compress,traffic, cmx, yuan3, yuan5, yuan6, traffic2, yuan8, yuan9}.
To address the needs of these diverse applications, previous research has developed a range of methods for calculating trajectory distances, including DTW~\cite{dtw,dtw_prun}, SSPD~\cite{sspd}, EDR~\cite{edr}, and Hausdorff~\cite{hausdorff}. However, these methods are inherently constrained by computational challenges due to their super-quadratic time complexity.
Therefore, recent advancements in trajectory similarity research have incorporated representation learning techniques to enhance the efficiency of similarity distance computations, as highlighted in several studies~\cite{TMN,trajCL, TrajGAT, road1, road2, road3}. These approaches first transform trajectories into high-dimensional embeddings in Euclidean space, and then use some distance functions between embeddings such as Euclidean distance to measure the 
trajectory similarity distance.

\begin{figure} % [h!] 帮助确定图片的精确位置
\vspace{-0.07in}
\centering % 使图片居中显示
\scalebox{0.8}{
\includegraphics[width=0.95\linewidth]{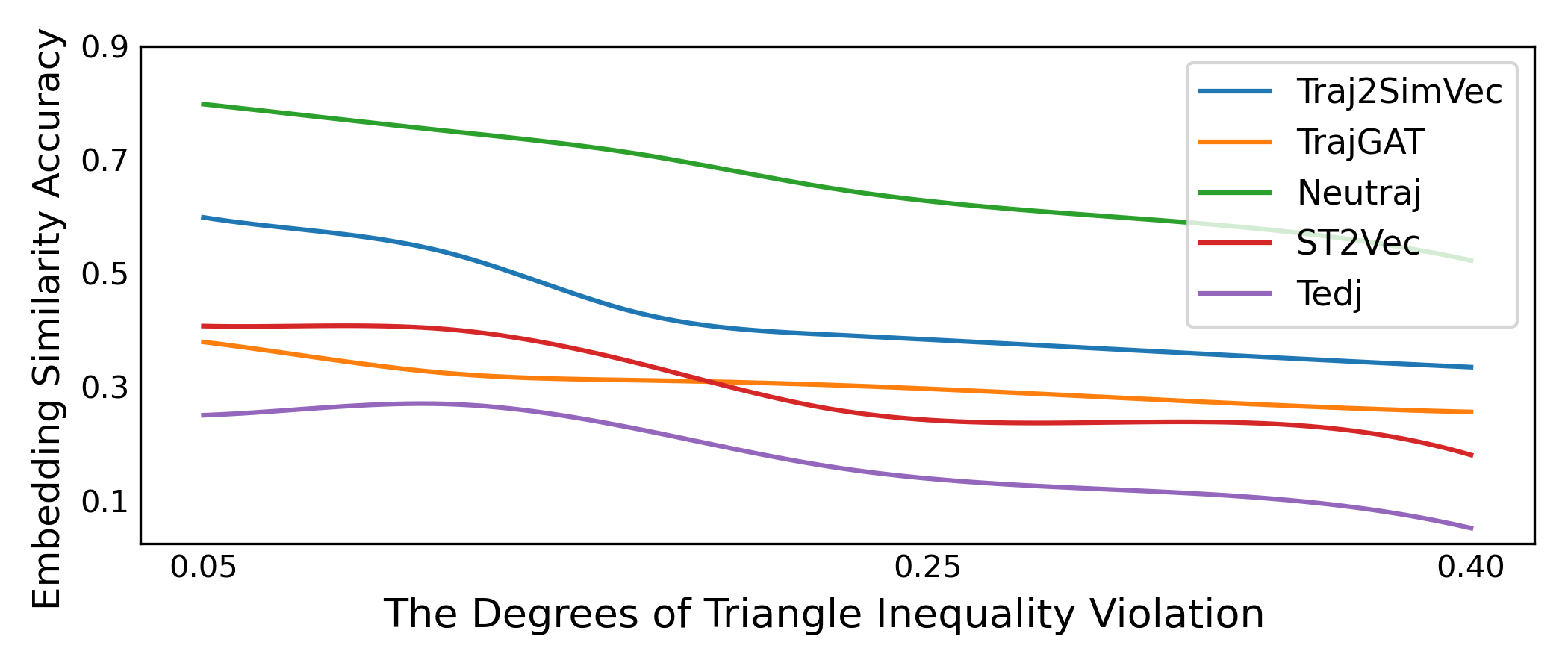}
% \caption{Relationship between the Strength of Triangle Inequality Contradiction and the Similarity Precision of Embedding}
}
\vspace{-0.15in}
\caption{\textcolor{black}{\small{Embedding Accuracy v.s. Triangle Inequality Violation}}}
 % 图片的标题
\label{fig:tri_str} % 用于文内引用的标签
\vspace{-0.3in}
\end{figure}

% \marginpar{\vspace{-3cm}\scriptsize{\textcolor{blue}{R2.O1}}}
However, this approach overlooks a critical issue: \textbf{\textit{While metrics between different embeddings in Euclidean space are constrained by the triangle inequality~\cite{eu,eu_metric_book,min_metric_book}, the similarities derived using traditional measures such as DTW and EDR on raw trajectories may not adhere to this constraint.}} As illustrated in Figure~\ref{fig:tri_str}, we evaluate the performance of existing advanced trajectory representation learning methods by sampling trajectories with varying degrees of triangular inequality violation\footnote{The triangular inequality violation among trajectories refers to the proportion of trajectory pairs that do not satisfy the triangle inequality. This is quantified by the average relative violation, as defined in Section~\ref{subsec:triag_vio}.}. The results indicate that the accuracy of these methods significantly deteriorates as the degree of triangular inequality violation increases. To overcome the limitations of Euclidean space in trajectory representation learning, we introduce the concept of non-Euclidean Hyperbolic space, which shows promise in bypassing these constraints~\cite{Colla, learningfeature, Learningcont}. Moreover, to effectively implement this in trajectory representation learning, we identify and address three critical challenges with corresponding solutions detailed in this paper.

\vspace{-0.06in}
\smallskip
\textbf{Challenge 1 (C1)}: \textit{Designing a suitable distance function for the hyperbolic space.} The ideal distance function should be non-negative, with a trajectory being most similar to itself, and should not be constrained by the triangle inequality.

\vspace{-0.02in}
\textbf{Challenge 2 (C2)}: \textit{ Effectively achieving hyperbolic embedding using current trajectory encoding models.} Directly transforming trajectories into hyperbolic space vectors is challenging, as these vectors must adhere to specific hyperbolic constraints that existing models cannot effectively handle. Therefore, developing a model-agnostic approach for hyperbolic embedding is essential, as it ensures compatibility with real-world applications and seamless integration with existing embedding frameworks without requiring modifications.
% It is crucial to develop a model-agnostic approach that allows hyperbolic embedding, which is especially suitable for real-world applications and integrable with existing embedding frameworks without modifications.

\textbf{Challenge 3 (C3)}: \textit{ Adapting to varied magnitudes of triangle inequality constraints among different trajectory pairs.} The extent of triangle inequality constraints varies across different similarity functions and datasets, necessitating a dynamic mechanism that ensures robust embeddings adaptable to diverse scenarios.

\smallskip
\noindent \textbf{Our proposed solution:}
\noindent First, to address \textbf{C1}, we design the Lorentz distance, calculated via the Lorentz inner product, to quantify the similarity between embeddings. We demonstrate that this distance is not bound by the triangle inequality, making it a robust measure of similarity.
\noindent Second, for \textbf{C2}, we introduce a model-agnostic \textbf{L}orentzian \textbf{H}yperbolic plugin (\textbf{LH}-plugin) that seamlessly integrates with any existing trajectory embedding model without necessitating alterations to the original frameworks. We first explore a vanilla hyperbolic projection method for converting Euclidean embeddings into hyperbolic ones. However, upon observing and proving that this projection can diminish distances as embedding norms increase, we refine our approach. In particular, we develop an enhanced projection technique using the hyperbolic cosine function, ensuring a more effective and theoretically sound mapping that avoids the issue of diminishing distances.
\noindent Lastly, to tackle \textbf{C3}, we assess the impact of varying triangle inequality constraints across datasets and similarity measures, and then introduce the Dynamic Fusion Distance, which adjusts to the unique characteristics of each trajectory pair. This approach dynamically blends Lorentzian and Euclidean distances, creating a hybrid metric tailored to the magnitude of triangle inequality constraints.

In summary, we make the following contributions:

\begin{itemize}[leftmargin=10.2pt]
\setlength{\itemsep}{0pt}
\setlength{\parsep}{0pt}
\setlength{\parskip}{0pt}
\item To the best of our knowledge,  we are the first to integrate hyperbolic space into trajectory representation learning.  We design the Lorentz distance, rigorously proving its ability to overcome triangle inequality constraints in this space. (Section~\ref{sec:pre})

\item We develop a model-agnostic framework \textbf{LH}-plugin that facilitates the integration of hyperbolic embeddings into the existing trajectory representation learning pipeline.  We provide a theoretical justification for this optimization. (Sections~\ref{sec:framework} and \ref{sec:projection})

\item We introduce a dynamic fusion method that adjusts the distance according to trajectory pair combinations. This innovative approach adapts to the diverse magnitudes of triangle inequality constraints encountered across trajectory pairs. (Section~\ref{sec:Fusion_dist})

\item We conduct comprehensive experiments with several state-of-the-art trajectory embedding models across various real-world datasets. Our results validate the effective integration of the \textbf{LH}-plugin into these models. (Section~\ref{sec:exp})
\end{itemize}

\section{Preliminary}
\label{sec:pre}

In this section, we first introduce the original pipeline of trajectory embedding in Euclidean space and discuss the triangle inequality problem encountered by this method for similarity distance calculation in Section \ref{sec:pre_traj_emb}. Next, we propose an innovative Lorentz distance between different embeddings within the hyperbolic space in Section~\ref{sec:pre_hyper_loren}, demonstrating its effectiveness as a distance measure while avoiding the triangle inequality constraint. Finally, we present the motivation behind our plugin-based approach for trajectory embedding in hyperbolic space in Section~\ref{sec:cmp_framework}.

% \marginpar{\vspace{-4cm}\scriptsize{\textcolor{blue}{R1.O4}}}
% \begin{longtable}{|c|c|c|c|c|}
% \hline
% \multicolumn{5}{|c|}{Symbol Table} \\
% \hline
% Symbol & Explanation & Symbol & Explanation \\
% \hline
% \endfirsthead
% \hline
% Symbol & Explanation & Symbol & Explanation \\
% \hline
% \endhead
% \hline
% \endfoot

% % Example rows (you can add as many rows as needed)
% $\alpha$ & Alpha symbol & $\beta$ & Beta symbol \\
% $\gamma$ & Gamma symbol & $\delta$ & Delta symbol \\
% $\lambda$ & Lambda symbol & $\mu$ & Mu symbol \\
% \end{longtable}

\begin{table}[t]
    \centering
    % \caption{Symbol Table}
    \setlength{\tabcolsep}{3pt}
    \begin{adjustbox}{center} 
    \resizebox{0.88\linewidth}{!}{
    % \scalebox{0.87}{
    \begin{tabular}{cc|cc}
    \toprule
    
    \multicolumn{4}{c}{\textcolor{black}{Symbol   Table}} \\ \hline
    \textcolor{black}{$T$, $T_a$} & Trajectory & \textcolor{black}{$\textbf{Dist}_*$} & Trajectory Similarity distance \\ 
    \textcolor{black}{$\textbf{x}$, $\textbf{x}_a$} & Trajectory Embedding    & \textcolor{black}{$\textbf{d}_{Eu}$} & Euclidean distance \\ 
    \textcolor{black}{$\mathcal{H}(\beta)$} & Hyperbolic space (shape $\beta$) & \textcolor{black}{$\textbf{d}_{Lo}$} & Lorentz distance \\ 
    \textcolor{black}{$\textbf{a}$} & Hyperbolic point & \textcolor{black}{$\langle\textbf{a}, \textbf{b}\rangle$} & Lorentz inner product\\ 
    \textcolor{black}{$\textbf{x}^\mathcal{E}$} & Euclidean Embedding 
    & \textcolor{black}{$\phi$} & Vanilla Hyperbolic Projection \\
    \textcolor{black}{$\textbf{x}^\mathcal{H}$} & Hyperbolic Embedding
    & \textcolor{black}{$\phi_{cosh}$} & Cosh Hyperbolic Projection \\
    \textcolor{black}{\textbf{TVF}}& Triangle Violation Flag& \textcolor{black}{\textbf{RV}} & Ration of Violation\\
    \textcolor{black}{\textbf{RVS}} & Relative Violation Scale & \textcolor{black}{$\alpha^{Lo}$} & Lorentz proportions\\
    \bottomrule
    \end{tabular}
    }
    \end{adjustbox}
    \vspace{-0.2in}
\end{table}

\vspace{-0.06in}
\subsection{Trajectory Embedding}
\vspace{-0.05in}
\label{sec:pre_traj_emb}

\noindent\textbf{Trajectory.} The trajectory $T = [p_1,...,p_n]$ is a sequence of points where each point $p_i = (lon_i,lat_i)$ (or $p_i= (lon_i,lat_i,t_i)$) is a tuple, and $lon_i, lat_i$ (and $t_i$) respectively represent longitude, latitude (and timestamp).

\noindent\textcolor{black}{
\textbf{Trajectory Similarity with Distance Measurement.} Given two trajectories $T_a=[p_{a_1},...,p_{a_n}]$ and $T_b=[p_{b_1},...,p_{b_m}]$, the similarity between $T_a$ and $T_b$ is measured by 
% the distance between the corresponding point sequences $[p_{a_1},...,p_{a_n}]$ and $[p_{b_1},...,p_{b_m}]$, measured using 
specific distance metrics on the corresponding point sequences, such as DTW~\cite{dtw}, Hausdorff~\cite{hausdorff}, SSPD~\cite{sspd}, EDR~\cite{edr}, etc. In particular, the higher the similarity, the lower the trajectory distance. For simplicity, we use $\mathbf{Dist}_*(\cdot,\cdot)$ to denote the trajectory distance.
% To avoid potential misunderstandings on this matter, we use \textbf{trajectory distance}, $\mathbf{Dist}_*(\cdot,\cdot)$, to describe this concept in the subsequent sections.
For instance, the DTW trajectory distance can be succinctly described as follows:
}
\vspace{-0.17in}

\begin{small}
\begin{equation}
\label{sim_dtw_equ}
\begin{aligned}
&\mathbf{Dist}_{D}(T_a, T_b) = DTW[p_{a_n},p_{b_m}]\\[-0.02in]
&DTW[p_{a_i},p_{b_j}] = d(p_{a_i},p_{b_j}) + \min \left(DTW[p_{a_{i-1}},p_{b_j}],\right.\\[-0.03in]
&\quad \quad \quad\quad \quad\quad\quad\quad \left.DTW[p_{a_i},p_{b_{j-1}}],DTW[p_{a_{i-1}},p_{b_{j-1}}]\right)\\[-0.08in]
\end{aligned}
\end{equation}
\end{small}

\noindent where $d(p_{a_i}, p_{b_j})$ represents the Euclidean distance between points $p_{a_i}$ and $ p_{b_j}$.  

\textit{Remark.} To prevent potential misunderstandings, we use \(\mathbf{Dist}_*(\cdot,\cdot)\) interchangeably to represent either trajectory similarity or distance in this paper.

\noindent\textbf{Trajectory Embedding.} 
The concept of trajectory embedding involves identifying an encoding function $g(\cdot)$ called the Trajectory Embedding model, which maps a trajectory onto a d-dimensional vector, represented as $\textbf{x}=g(T)\in \mathbb{R}^d$. 

\noindent\textbf{Trajectory Embedding for Similarity Computation.}
The primary objective of trajectory similarity embedding is to develop an encoder model such that the distance of embedding pair $(\textbf{x}_a,\textbf{x}_b)$, for any trajectory pair $(T_a, T_b)$, closely approximates the provided trajectory distance measure, such as $\mathbf{Dist}_D(T_a, T_b)$.
Prior studies have treated the corresponding trajectory embeddings $\textbf{x}_a$ and $\textbf{x}_b$ as vectors within Euclidean space, employing Euclidean distance to gauge the distance between these vectors. Given a database $\mathcal{T}=\{T_1, T_2,...,T_N \}$ comprising $N$ trajectories, alongside a distance measurement function $\mathbf{Dist}_*(\cdot, \cdot)$, the aim of trajectory similarity embedding is to discover a mapping $g(\cdot)$ that reduces the disparity between the Euclidean distance among trajectory embeddings 
$\mathbf{d}_{Eu}(g(T_i), g(T_j))$
and the actual distance $\mathbf{Dist}_*(T_i, T_j)$ to the minimum. This process can be depicted as follows:
\vspace{-0.22in}

\begin{equation}
\small
\label{eu_embedding}
       \underset{g}{\textbf{argmin}} \sum_{T_i, T_j \in \mathcal{T}} \Big|\ \mathbf{Dist}_*(T_i,T_j) - \mathbf{d}_{Eu}(g(T_i), g(T_j))\ \Big|\\[-0.1in]
\end{equation}

\noindent\textbf{Triangle Inequality in Trajectory Embedding.}
\label{sec:pre_traj_ineq}
Because the trajectory embedding space is a d-dimensional Euclidean space, the embedding distance between different trajectories satisfies the triangle inequality constraint as follows:
\vspace{-0.05in}
\begin{equation}
\small
\label{triangle_inequality1}
\begin{aligned}
    \forall T_a, T_b, T_c \in \mathcal{T},\quad \textbf{x}_*&=g(T_*)\\[-0.03in]
    \textbf{s.t.}\ \mathbf{d}_{Eu}(\textbf{x}_a,\textbf{x}_b)+\mathbf{d}_{Eu}(\textbf{x}_b,\textbf{x}_c) &\geq 
    \mathbf{d}_{Eu}(\textbf{x}_a,\textbf{x}_c)\\[-0.05in]
\end{aligned}
\end{equation}

\noindent\textbf{Triangle Inequality Violation in Trajectory Similarity.} Many trajectory distances, such as DTW and SSPD, don't satisfy the triangle inequality. Specifically, we can use a specific example to clarify this issue as follows.

% However, this constraint contradicts certain actual trajectory distances mentioned above, such as DTW, SSPD, and so on, which don't satisfy the triangle inequality. Specifically, we can use a specific example to clarify this issue as follows.

\begin{figure} % [h!] 帮助确定图片的精确位置
\centering % 使图片居中显示
\vspace{-0.05in}
\includegraphics[width=0.72\linewidth]{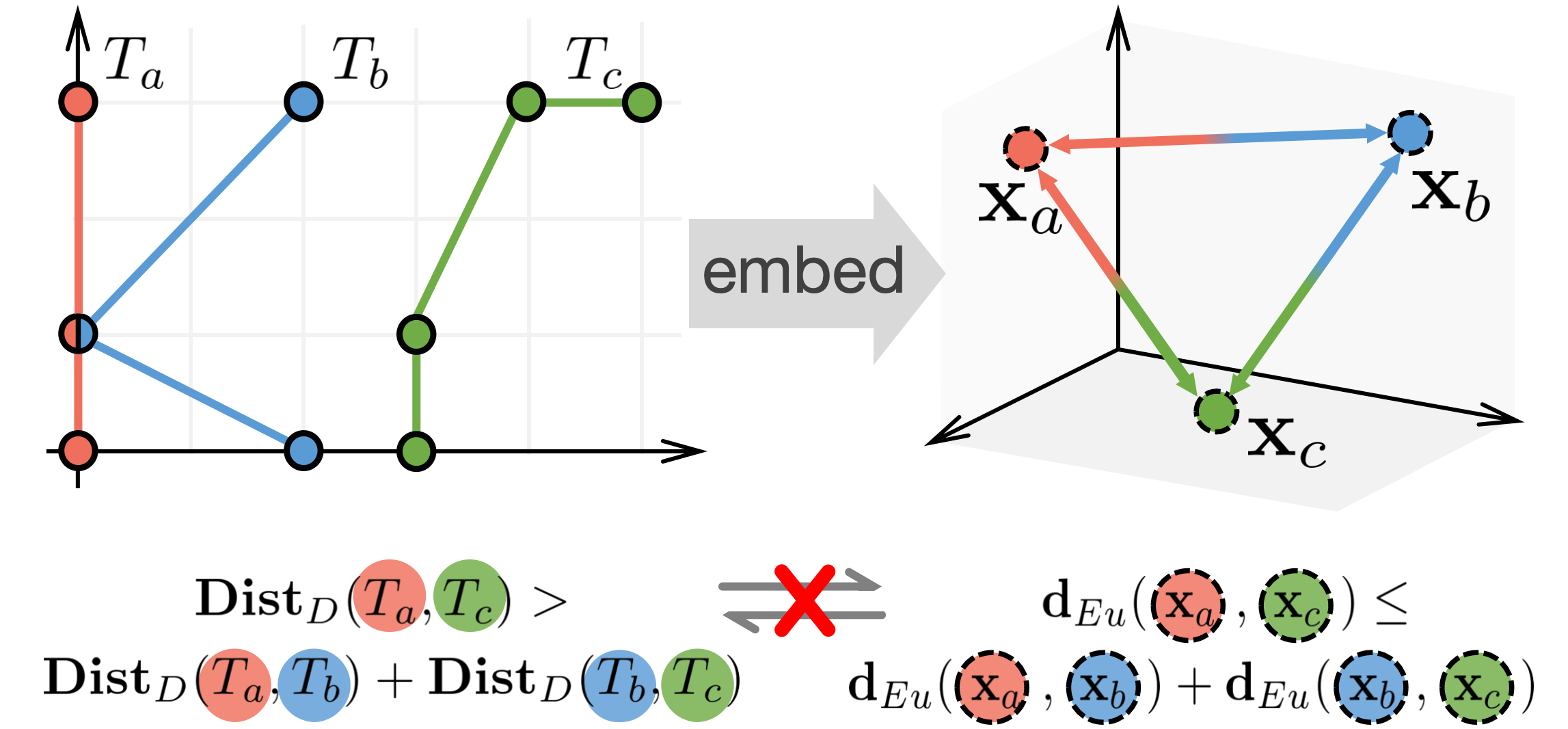}
\vspace{-0.08in}
\caption{An Example of Triangle Inequality Contradiction} % 图片的标题
\vspace{-0.28in}
\label{fig:example} % 用于文内引用的标签
\end{figure}

\vspace{-0.1in}
\begin{example}
% \small
In Figure~\ref{fig:example}, let $T_a=[(0,0),(0,1),(0,3)]$, $T_b=[(2,0),$ $(0,1),$ $(2,3)]$ and $T_c=[(3,0),(3,1),$ $(4,3),(5,3)]$. Using the DTW distance of Formula \ref{sim_dtw_equ}, the calculated trajectory distances are $\mathbf{Dist}_{D}(T_a, T_b) = 4$, $\mathbf{Dist}_{D}(T_b, T_c) = 9$, and $\mathbf{Dist}_{D}(T_a, T_c) = 15$. Notably, $\mathbf{Dist}_{D}(T_a, T_c) > \mathbf{Dist}_{D}(T_a, T_b) + \mathbf{Dist}_{D}(T_b, T_c)$, which is contradicted with the constraint in Formula~\ref{triangle_inequality1}. 
\end{example}

\vspace{-0.08in}
Consequently, for most similarity methods that do not satisfy the triangle inequality, the approximations of similarity through representation learning might inevitably deviate from the true similarity values. In certain cases, this discrepancy could even amplify, leading to distortions in the trajectory embeddings.

\vspace{-0.09in}
\subsection{Hyperbolic Space and Lorentz Distance}
\label{sec:pre_hyper_loren}
\vspace{-0.05in}

% \marginpar{\vspace{9cm}\scriptsize{\textcolor{blue}{R3.D3}}}
\noindent \textbf{Hyperbolic Space.} To avoid the triangle inequality constraint of high-dimensional embeddings in Euclidean space\footnote{\textcolor{black}{In Euclidean space, the metric must satisfy four conditions, one of which is the triangle inequality. In other words, Euclidean distance is already the metric with the fewest constraints in Euclidean space.}}, we introduce the non-Euclidean hyperbolic space, which has the potential to bypass this limitation~\cite{Colla, learningfeature, Learningcont}. 
In particular, this hyperbolic space is built on the Lorentz inner product, denoted as $\langle\textbf{a}, \textbf{b}\rangle = -a_0b_0 + \sum_{i=1}^{n}a_ib_i$, where $\textbf{a}=[a_0,\cdots,a_{n}]\in \mathbb{R}^{n+1} $ and $\textbf{b}=[b_0,\cdots,b_{n}]\in \mathbb{R}^{n+1} $ are two $n+1$-dimensional vectors. This space consists of all high-dimensional vectors that satisfy a specific condition within the Lorentz inner product, as described below.

\vspace{-0.08in}
\begin{definition}[Hyperbolic Space]
\textcolor{white}{23}

\textcolor{white}{123}

\vspace{-0.4in}

\begin{equation} 
\vspace{-0.02in}
\label{hyper_space}
\small
  \mathcal{H}(\beta) = \{\mathbf{a}\in\mathbb{R}^{n+1}:\ \langle\mathbf{a},\mathbf{a}\rangle = -\beta,\ a_0 \geq \sqrt{\beta}\} \quad (\beta >0)\\[-0.05in]
\end{equation}
where $ \langle\mathbf{a},\mathbf{a}\rangle$ represents the self Lorentz inner product, and $\beta$ is a hyper-parameter that influences the shape of the space.
\end{definition}
\vspace{-0.06in}

\noindent \textbf{Lorentz Distance.} Given a hyperbolic space $\mathcal{H}(\beta)$, we define the function \textit{Lorentz Distance} based on the Lorentz inner product as follows:
\vspace{-0.06in}

\begin{definition}[Lorentz Distance]
\textcolor{white}{23}

\textcolor{white}{123}

\vspace{-0.32in}
\begin{equation}
\label{lorenz_dist}
\small
  \mathbf{d}_{Lo}(\mathbf{a},\mathbf{b}) = |\langle\mathbf{a},\mathbf{b}\rangle| -\beta \qquad (\mathbf{a},\mathbf{b} \in \mathcal{H}(\beta))\\[-0.1in]
\end{equation}
where $|\langle\mathbf{a},\mathbf{b}\rangle|$ means the absolute value of $\langle\mathbf{a},\mathbf{b}\rangle$.
\end{definition}
\vspace{-0.07in}

% We can use the function \textit{Lorentz Distance} to measure the distance between any two embeddings in the space due to its non-negativity and minimal value of self-distance as described in Lemma~\ref{lem:distance}. Additionally, it can relax the constraint of the triangle inequality as explained in Lemma~\ref{lem:triangle}.
\ifnum\tech=0
We can use the function \textit{Lorentz Distance} to measure the distance between any two embeddings in the space due to its non-negativity and minimal value of self-distance as described in Lemma~\ref{lem:distance}. Additionally, it can relax the constraint of the triangle inequality as explained in Lemma~\ref{lem:triangle}. 

% \textcolor{red}{Due to space limitations, the details of the complex proof process for these two Lemmas are provided in the technical report~\cite{Tech_Rpt}.}
\else
We can use the function \textit{Lorentz Distance} to measure the distance between any two embeddings in the space due to its non-negativity and minimal value of self-distance as described in Lemma~\ref{lem:distance}. Additionally, it can relax the constraint of the triangle inequality as explained in Lemma~\ref{lem:triangle}.
\fi

\vspace{-0.06in}

\begin{lemma}
\label{lem:distance}

$\forall\mathbf{a},\mathbf{b}\in \mathcal{H}(\beta)$, $\mathbf{d}_{Lo}(\mathbf{a},\mathbf{b})\geq 0$, and $\mathbf{d}_{Lo}(\mathbf{a},\mathbf{b})=0$ only when $\mathbf{a}=\mathbf{b}$.
\vspace{-0.09in}

\ifnum\tech=0 

\else

\begin{prf}
\small
\setstretch{1.15}
Suppose $\textbf{a}=[a_0,\cdots,a_n] $ and $\textbf{b}=[b_0,\cdots, b_n]$ are any two embeddings in the space $\mathcal{H}(\beta)$. We have $a_0 \geq \sqrt{\beta}$, $b_0 \geq \sqrt{\beta}$, $a_{0}^{2} - \sum_{i=1}^{n} a_{i}^{2} = \beta$, and $b_{0}^{2} - \sum_{i=1}^{n} b_{i}^{2} = \beta$. Define a function $f(x) = \beta x^2 - 2\langle \textbf{a}, \textbf{b} \rangle x + \beta$. We can further express this function by extending the Lorentz inner product and replacing $\beta$ in the quadratic coefficient and constant term with $a_{0}^{2} - \sum_{i=1}^{n} a_{i}^{2}$ and $b_{0}^{2} - \sum_{i=1}^{n} b_{i}^{2}$, respectively, as follows:
\vspace{-0.25in}

\begin{equation}
\small
\begin{aligned}
    f(x) &= (a_0^2 - \sum_{i=1}^{n} a_i^2)x^2 + 2(a_0 b_0 - \sum_{i=1}^{n} a_i b_i)x + (b_0^2 - \sum_{i=1}^{n} b_i^2) \\[-0.13in]
         &= (a_0 x + b_0)^2 - \sum_{i=1}^{n} (a_i x + b_i)^2\\[-0.12in]
\end{aligned}
\end{equation}
Let $x = -\frac{b_0}{a_0} < 0$. We have:
\vspace{-0.12in}
\begin{equation}
\small
    f\left(-\frac{b_0}{a_0}\right) = - \sum_{i=1}^{n} \left(-a_i \frac{b_0}{a_0} + b_i\right)^2 \leq 0\\[-0.08in]
\end{equation}
Since the quadratic coefficient $\beta > 0$, we have $f(-\infty) > 0$ and $f(+\infty) > 0$. Therefore, the quadratic equation $f(x) = 0$ has at least one root. The discriminant $\Delta$ of $f(x) = 0$ is given by:
\vspace{-0.11in}

\begin{equation}
    \Delta = (2\langle \textbf{a}, \textbf{b} \rangle)^2 - 4 \beta^2 \geq 0 \\[-0.06in]
\end{equation}
Thus, we have $\langle \textbf{a}, \textbf{b} \rangle^2 \geq \beta^2 \Rightarrow |\langle \textbf{a}, \textbf{b} \rangle| \geq \beta \Rightarrow \textbf{d}_{Lo}(\textbf{a}, \textbf{b}) = |\langle \textbf{a}, \textbf{b} \rangle| - \beta \geq 0$.
In particular, if $\textbf{d}_{Lo}(\textbf{a}, \textbf{b}) = |\langle \textbf{a}, \textbf{b} \rangle| - \beta = 0$, we have $f(x)\geq 0$ and only one root for the quadratic equation $f(x) = 0$. Considering that $f(-\frac{b_0}{a_0})\leq 0$, the root should be $x = -\frac{b_0}{a_0} < 0$. The root can also be computed as $x = \frac{-(-2 \langle \textbf{a}, \textbf{b} \rangle)}{2\beta} = \frac{\langle \textbf{a}, \textbf{b} \rangle}{\beta}$ according to the root-finding algorithm. Given that $|\langle \textbf{a}, \textbf{b} \rangle| - \beta = 0$ and $x < 0$, we find the root $x = -1$. Therefore, $x = -\frac{b_0}{a_0} = -1 \Rightarrow a_0 = b_0$. Additionally:

\vspace{-0.3in}
\begin{equation}
    f(-1) = (b_0 - a_0)^2 - \sum_{i=1}^{n} (b_i - a_i)^2 = -\sum_{i=1}^{n} (b_i - a_i)^2 = 0\\[-0.1in]
\end{equation}
Thus, $b_i - a_i = 0$ is satisfied for each $i \in [0, n]$, implying $\textbf{b} = \textbf{a}$.
\vspace{-0.2in}
\end{prf}

\fi

\end{lemma}

\begin{lemma}\label{lem:triangle}
$\exists \mathbf{a},\mathbf{b}, \mathbf{c} \in \mathcal{H}(\beta)$, the triangle inequality $\mathbf{d}_{Lo}(\mathbf{a},\mathbf{b})\leq  \mathbf{d}_{Lo}(\mathbf{a},\mathbf{c}) + \mathbf{d}_{Lo}(\mathbf{b},\mathbf{c})$ is not satisfied.
\vspace{-0.06in}

\ifnum\tech=0 

\else

\begin{prf}
\small
\setstretch{1.15}
We need to find at least one set of three embeddings \(\mathbf{a}, \mathbf{b}, \mathbf{c}\) that does not satisfy this inequality. Given any \(a_0 > \sqrt{\beta}\), consider its corresponding embedding \(\mathbf{a} = [a_0, a_1, \ldots, a_n] \in \mathcal{H}(\beta)\). Thus, we have \(\langle \mathbf{a}, \mathbf{a} \rangle = a_0^2 - \sum_{i=1}^n a_i^2 = -\beta\).
Next, consider another embedding \(\mathbf{b} = [a_0, -a_1, \ldots, -a_n]\). We also have \(\langle \mathbf{b}, \mathbf{b} \rangle = a_0^2 - \sum_{i=1}^n (-a_i)^2 = -\beta\). Therefore, \(\mathbf{b}\) also belongs to \(\mathcal{H}(\beta)\).
Finally, consider the embedding \(\mathbf{c} = [\sqrt{\beta}, 0, \ldots, 0]\), which also belongs to the space due to \(\langle \mathbf{c}, \mathbf{c} \rangle = -(\sqrt{\beta} \cdot \sqrt{\beta}) = -\beta\).
Now, we compute the distances between the three embeddings:
\vspace{-0.3in}

\[
\begin{aligned}
\mathbf{d}_{Lo}(\mathbf{a}, \mathbf{b}) &= |\langle \mathbf{a}, \mathbf{b} \rangle| - \beta = |-a_0^2 + \sum_{i=1}^n -a_i^2| - \beta \\[-0.16in]
&= |-2a_0^2 + (a_0^2-\sum_{i=1}^n a_i^2)| - \beta = 2a_0^2 - 2\beta\\[-0.07in]
\mathbf{d}_{Lo}(\mathbf{a}, \mathbf{c}) &= |\langle \mathbf{a}, \mathbf{c} \rangle| - \beta = |-a_0 \sqrt{\beta}| - \beta = a_0 \sqrt{\beta} - \beta\\[-0.02in]
\mathbf{d}_{Lo}(\mathbf{b}, \mathbf{c}) &= |\langle \mathbf{b}, \mathbf{c} \rangle| - \beta = |-a_0 \sqrt{\beta}| - \beta = a_0 \sqrt{\beta} - \beta
\end{aligned}\]
Therefore, we have:
\vspace{-0.1in}
\[
\mathbf{d}_{Lo}(\mathbf{a}, \mathbf{c}) + \mathbf{d}_{Lo}(\mathbf{b}, \mathbf{c}) = 2a_0 \sqrt{\beta} - 2\beta\\[-0.1in]
\]
Comparing this with \(\mathbf{d}_{Lo}(\mathbf{a}, \mathbf{b})\):
\vspace{-0.15in}

\[
2a_0 \sqrt{\beta} - 2\beta < 2a_0^2 - 2\beta = \mathbf{d}_{Lo}(\mathbf{a}, \mathbf{b})\\[-0.1in]
\]
Thus, we have found three embeddings in the space that do not satisfy the triangle inequality.
\vspace{-0.19in}

\end{prf}
\fi

\end{lemma}

\ifnum\tech=0 

Due to space limit, the detailed proof process for these Lemmas is elaborated in the accompanying technical report~\cite{Tech_Rpt}.

\fi

\vspace{-0.08in}
\subsection{Embedding Methods in Hyperbolic Space}
\label{sec:cmp_framework}
\vspace{-0.05in}

The intuitive method for mapping trajectories into hyperbolic space is to directly transform operators from Euclidean to hyperbolic space, constructing hyperbolic neural network layers. For instance, Ganea et al.~\cite{hnn} adapt structures like MLP and CNN from Euclidean space to hyperbolic neural networks by combining Möbius gyrovector spaces with the Riemannian geometry of the Poincaré model. This allows the entire trajectory embedding process to operate in hyperbolic space, requiring the model to be built from scratch. However, this method has two major drawbacks: 1) Current GPU optimizations are designed for standard neural network operators in Euclidean space, and there is no comparable optimization for hyperbolic neural network operators. This results in significantly reduced efficiency for training and inference in purely hyperbolic models. 2) Adapting existing trajectory embedding methods to pure hyperbolic transformations is highly complex. Aside from neural networks, there are no standard hyperbolic space solutions for data processing and training in these specific embedding methods.

In contrast, we introduce the \textit{Plugin-based} approach in this paper. This method retains as much of the Euclidean space components in trajectory embedding as possible, only introducing hyperbolic space in the output phase by mapping Euclidean vectors to hyperbolic ones. This approach preserves the original embedding pipeline, fully utilizing the efficiency of neural networks in Euclidean space while maintaining the techniques used in existing embedding methods. As a result, it enables a model-agnostic transition to hyperbolic embedding without sacrificing speed or flexibility.

\vspace{-0.05in}
\section{Framework Overview}
\label{sec:framework}
\vspace{-0.04in}

\noindent \textbf{Motivation.} Existing trajectory embedding works, such as those described in~\cite{neutraj, TrajGAT, traj2s, st2vec}, typically assess trajectory similarity using Euclidean distance between embeddings, a measure constrained by the triangle inequality. As explored in Section~\ref{sec:pre_hyper_loren}, adopting hyperbolic space and the Lorentz distance can mitigate these limitations. Accordingly, our goal is to enhance trajectory embeddings by incorporating these insights. Rather than developing an entirely new embedding model, we introduce a lightweight framework named \textbf{LH}-plugin. This framework is designed to adapt existing embedding models, providing benefits in both training and inference efficiency, as discussed in Section~\ref{sec:cmp_framework}. Furthermore, the plugin-based approach ensures plug-and-play compatibility, avoiding disruptions to the existing model’s workflow and enhancing its practical robustness.

\begin{figure} % [h!] 帮助确定图片的精确位置
\vspace{-0.06in}
\centering % 使图片居中显示
\includegraphics[width=0.91\linewidth]{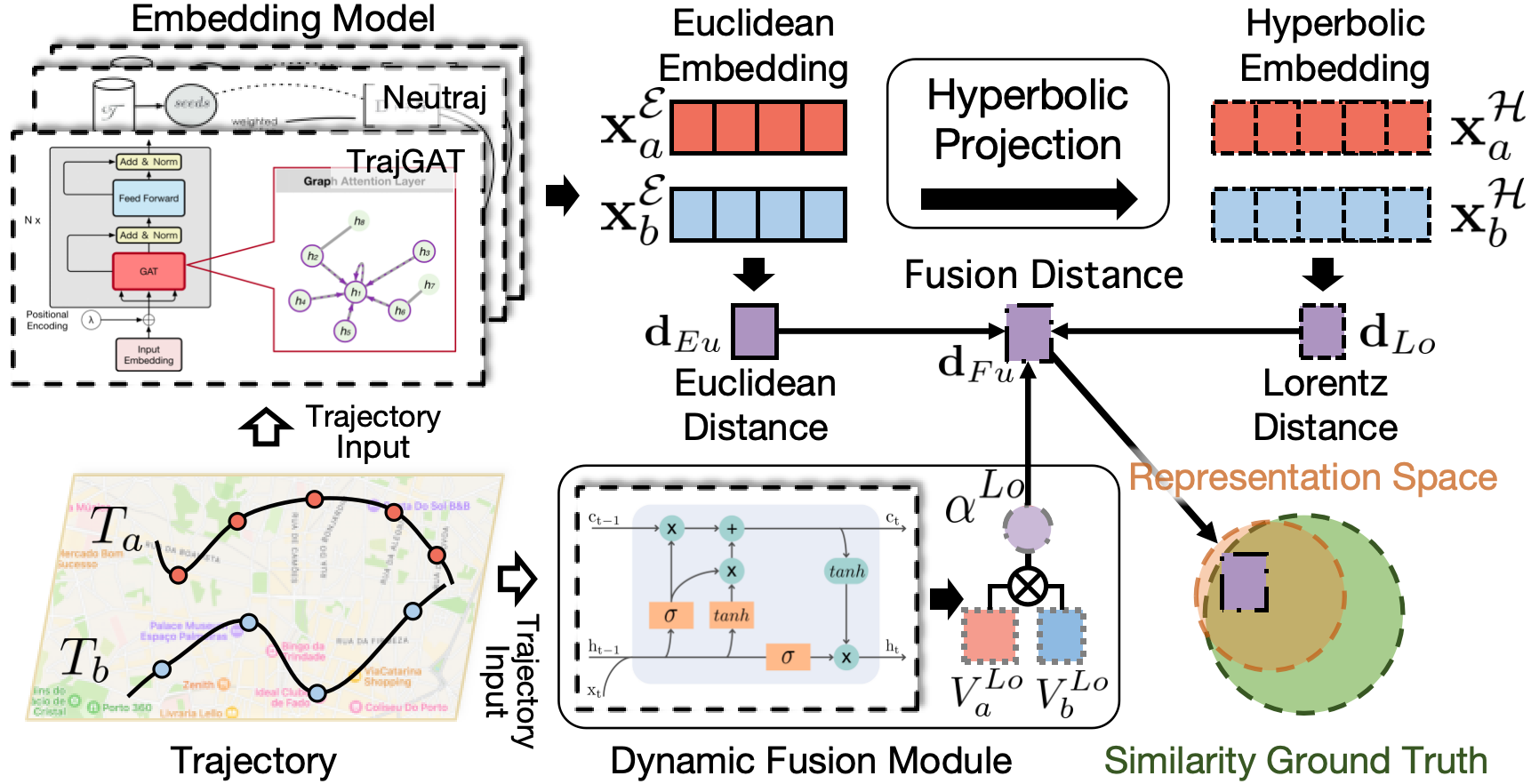} % 设置图片宽度为文本宽度的80%
\vspace{-0.05in}
\caption{\textcolor{black}{The Framework Overview of \textbf{LH}-plugin}} % 图片的标题
\label{fig:pipeline} % 用于文内引用的标签
\vspace{-0.28in}
\end{figure}
% Existing trajectory embedding works, such as~\cite{neutraj, TrajGAT, traj2s, st2vec}, process raw trajectories and generate corresponding embeddings in Euclidean space. These methods approximate trajectory similarity using the Euclidean distance between embeddings, inherently constrained by the triangle inequality. However, as discussed in Section~\ref{sec:pre_hyper_loren}, leveraging hyperbolic space and the Lorentz distance can alleviate these constraints. Consequently, we aim to enhance trajectory embeddings by integrating these observations. Instead of developing a new embedding model, we propose a lightweight framework named \textbf{LH}-plugin, designed to adapt existing embedding models. The reason, as discussed in Section\ref{sec:cmp_framework}, is that the plugin-based approach not only offers advantages in training and inference efficiency, but also ensures plug-and-play compatibility without altering the existing model's workflow, making it robust in practical scenarios.

% \marginpar{\vspace{-3cm}\scriptsize{\textcolor{blue}{R1.O1\\R3.D6}}}
As in Figure~\ref{fig:pipeline},  the framework incorporates two additional modules into the existing embedding pipeline: the \textit{Hyperbolic Projection} module, which maps embeddings into hyperbolic space, and the \textit{Dynamic Fusion} module, which dynamically fuses the Euclidean distance and Lorentz distance with a learnable ratio to measure trajectory similarity. Below, we explain how these modules enhance the existing similarity/distance computation workflow with trajectory embeddings. 

\noindent \underline{1. Generating Embeddings in Euclidean Space.} Given two trajectories \(T_a\) and \(T_b\), we first use existing embedding models such as \textbf{Neutraj} and \textbf{TrajGAT} to generate embeddings, denoted as \(\textbf{x}_a^\mathcal{E}\) and \(\textbf{x}_b^\mathcal{E}\), which are vectors in Euclidean space.

\noindent \underline{2. Mapping Embeddings to Hyperbolic Space.} We design a projection algorithm in the \textit{Hyperbolic Projection} module to transform \(\textbf{x}_a^\mathcal{E}\) and \(\textbf{x}_b^\mathcal{E}\) into hyperbolic embeddings, \(\textbf{x}_a^\mathcal{H}\) and \(\textbf{x}_b^\mathcal{H}\).

\noindent \underline{3. Computing Fusion Distance.} While hyperbolic embeddings mitigate triangle inequality constraints, there are instances where Euclidean distance is more appropriate. To leverage the strengths of both distances, we fuse them rather than relying solely on Lorentz distance. The fusion ratio varies across trajectory pairs, necessitating the \textit{Dynamic Fusion} module. This module generates a ratio $\alpha^{Lo}$ for the given trajectory input (\(T_a\) and \(T_b\), \textcolor{black} {that is, the white arrows in the figure}) 
and uses it to fuse the Euclidean distance \(\textbf{d}_{Eu}\) and hyperbolic distance \(\textbf{d}_{Lo}\) into the fusion distance \(\textbf{d}_{Fu}\). The \(\textbf{d}_{Fu}\) and the ground truth distance are then used to compute the loss for training models within the \textit{Dynamic Fusion} module.
% \marginpar{\vspace{-2.1cm}\scriptsize{\textcolor{blue}{R1.O1}}}

\vspace{-0.04in}
\section{Hyperbolic Projection}
\label{sec:projection}
\vspace{-0.03in}

In this section, we introduce the \textit{Vanilla Projection} method and analyze the challenges related to distance degradation in detail in Section~\ref{sec:projection_drawback}. To address these shortcomings, we subsequently introduce the \textit{Cosh Projection} method in Section~\ref{sec:projection_cosh}, which is designed to mitigate the identified issues effectively.

\vspace{-0.07in}
\subsection{Vanilla Hyperbolic Projection}
\label{sec:projection_drawback}
\vspace{-0.05in}
In an $n$-dimensional Euclidean space, an embedding $\textbf{x}^{\mathcal{E}} \in \mathbf{R}^n$ possesses $n$ degrees of freedom. In contrast, embeddings in an $n$-dimensional hyperbolic space exhibit only $n-1$ degrees of freedom due to an inherent constraint imposed by the space's definition. To align the degrees of freedom between these spaces, it is necessary to map an $n$-dimensional Euclidean embedding into a $(n+1)$-dimensional hyperbolic embedding $\textbf{x}^{\mathcal{H}} \in \mathbf{R}^{n+1}$. Specifically, each point within the Euclidean space $\mathcal{E}$ is uniquely mapped to a corresponding point in the hyperbolic space $\mathcal{H}$, thereby establishing a bijective relationship between the two spaces. This property is crucial in representation learning, ensuring that distinct trajectories $T$ and $T'$, represented by different Euclidean vectors $\textbf{x}^\mathcal{E}$ and $\textbf{x}'^{\mathcal{E}}$, do not converge to the same hyperbolic point $\textbf{x}^\mathcal{H}$ through the mapping function $\phi$. Failure to maintain such distinct mappings would result in trajectory confusion, leading to erroneous similarity assessments among trajectories.  To achieve this bijective mapping, a straightforward approach involves a direct projection technique. This technique extends points from the Euclidean space $\mathcal{E}$ into the hyperbolic space by introducing an additional dimension. Dimensions $[1, n]$ are preserved as-is, while the new dimension's value is determined to satisfy the hyperbolic plane $\mathcal{H}(\beta)$ conditions. We refer to this mapping strategy as the \textit{Vanilla Hyperbolic Projection} as follows:
% \marginpar{\vspace{-11cm}\scriptsize{\textcolor{blue}{R1.O4\\R3.D7}}}
\vspace{-0.3in}

\begin{equation*}
\small
\phi(\textbf{x}^\mathcal{E}) \rightarrow \textbf{x}^\mathcal{H}:
\ x_i^{\mathcal{H}} = 
\left\{
\begin{array}{ll}
x_{i-1}^{\mathcal{E}} & \text{if } 1 \leq i \leq n, \\
\sqrt{\sum_{j=1}^n (x_j^{\mathcal{E}})^2 + \beta} & \text{if } i = 0.
\end{array}
\right.
\vspace{-0.05in}
\end{equation*}

\begin{figure}[!t]
\vspace{-0.11in}
    \centering
    \scalebox{0.72}{
    \captionsetup{skip=3pt} 
    \begin{subfigure}[t]{0.45\linewidth} 
        \centering
        \includegraphics[height=4cm,keepaspectratio]{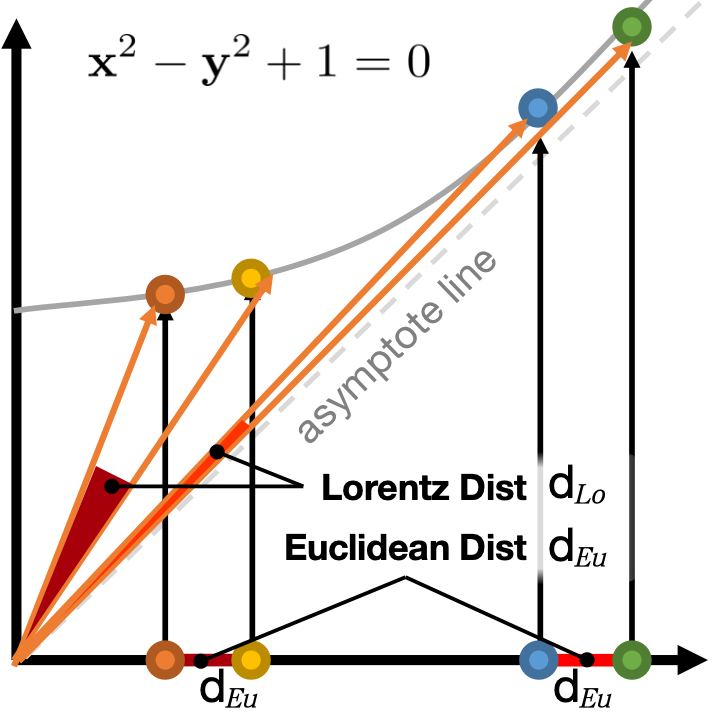}
        \vspace{-0.23in}
        \caption{The Vanilla Projection}
        \label{fig:mapping_err}
    \end{subfigure}%
    \hfill 
    \begin{subfigure}[t]{0.45\linewidth}
        \centering
        \includegraphics[height=4cm, keepaspectratio]{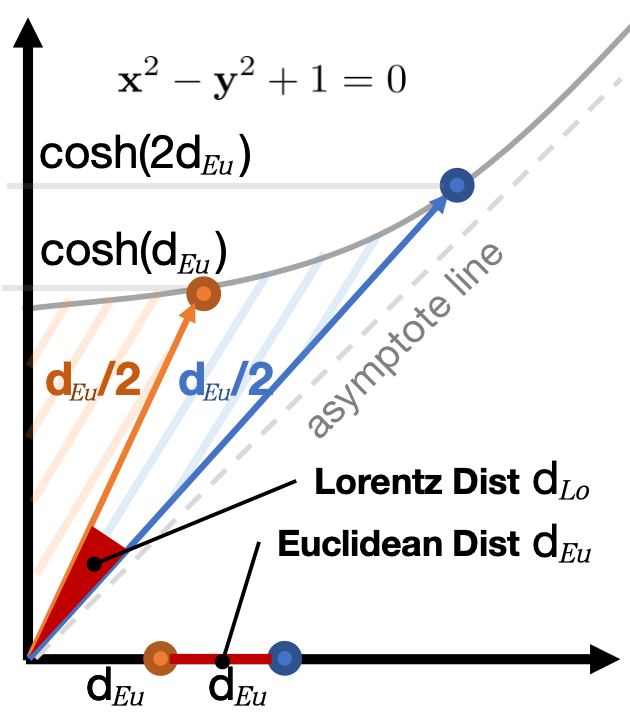}
        \vspace{-0.06in}
        \caption{The Cosh  Projection}
        \label{fig:cosh_mapping}
    \end{subfigure}
    }
    \vspace{-0.07in}
    \caption{\textcolor{black}{The comparison between Vanilla Hyperbolic Projection and Cosh Hyperbolic Projection}}
    \vspace{-0.29in}
\end{figure}

\noindent \textbf{Lorentz distance degradation analysis.}  However, a significant limitation of the Vanilla Hyperbolic Projection is its failure to preserve distances between different embeddings measured by the Lorentz distance, particularly when the magnitudes of the original embeddings are substantial. We employ a two-dimensional hyperbolic space for illustration to elucidate this issue more clearly. As depicted in Figure~\ref{fig:mapping_err}, we simplify the hyperbolic space to the equation $x^2 - y^2 + 1 = 0$, represented by the gray hyperbola. We consider points in $\mathcal{E}$ positioned along the positive x-axis. The black arrows in the figure demonstrate the projection process, while the orange arrows indicate the resultant vectors in the hyperbolic space. The red cones between the two vectors represent the angle between them. Significantly, the size of this angle exhibits a positive correlation with the Lorentz distance, particularly when the magnitudes of the Euclidean embeddings are large. 
This observation inspires us to deduce a theorem as follows.

\vspace{-0.05in}
\begin{theorem}\label{lem:mapping1}
For any two vectors \(\mathbf{a}=(a_0,\cdots,a_n)\) and \(\mathbf{b}=(b_0,\cdots,b_n)\) in \(\mathcal{H}(\beta)\), if \(a_0, b_0 \gg \beta\) and \(a_0, b_0 \gg |a_0 - b_0|\), then the Lorentz distance \(\mathbf{d}_{Lo}(\mathbf{a}, \mathbf{b}) \rightarrow 0\).

\vspace{0.02in}
\begin{prf}
\small
First, consider the unit vector \(\mathbf{e} = (1,0,\cdots,0)\) along the \(x\)-axis. We can compute the angles between the unit vector and \(\mathbf{a}\), \(\mathbf{b}\) as follows:

\vspace{-0.23in}
\[
\mathbf{a}, \mathbf{b} \in \mathcal{H} \Rightarrow a_0^2 - \sum_{i=1}^n a_i^2 = \beta, \quad b_0^2 - \sum_{i=1}^n b_i^2 = \beta \\[-0.20in]
\]

\[
\cos(\mathbf{a}, \mathbf{e}) = \frac{\mathbf{a} \cdot \mathbf{e}}{|\mathbf{a}| \cdot |\mathbf{e}|} = \frac{a_0}{\sqrt{a_0^2 + \sum_{i=1}^n a_i^2}} = \frac{a_0}{\sqrt{2a_0^2 - \beta}}\\[-0.17in]
\]
\[\\[-0.01in]\]
\[
\cos(\mathbf{b}, \mathbf{e}) = \frac{\mathbf{b} \cdot \mathbf{e}}{|\mathbf{b}| \cdot |\mathbf{e}|} = \frac{b_0}{\sqrt{b_0^2 + \sum_{i=1}^n b_i^2}} = \frac{b_0}{\sqrt{2b_0^2 - \beta}}\\[-0.1in]
\]
Given \(a_0, b_0 \gg \beta\), we have:
\vspace{-0.08in}
\[
\cos(\mathbf{a}, \mathbf{e}) \approx \cos(\mathbf{b}, \mathbf{e}) \approx \sqrt{2}/{2}\\[-0.05in]
\]

Thus, the angle between \(\mathbf{a}\) and \(\mathbf{b}\) is near zero, implying \(\cos(\mathbf{a}, \mathbf{b}) = 1\). Next, we compute the Lorentz distance between \(\mathbf{a}\) and \(\mathbf{b}\):
\vspace{-0.12in}
{\footnotesize
\[
\langle \mathbf{a}, \mathbf{b} \rangle = -a_0 b_0 + \sum_{i=1}^n a_i b_i = -2a_0 b_0 + \mathbf{a} \cdot \mathbf{b} = -2a_0 b_0 + |\mathbf{a}||\mathbf{b}|\cos(\mathbf{a}, \mathbf{b})\\[-0.09in]
\]}
Given \(\cos(\mathbf{a}, \mathbf{b}) \approx 1\), we get:
\vspace{-0.07in}
{\footnotesize
\[
\langle \mathbf{a}, \mathbf{b} \rangle \approx -2a_0 b_0 + \sqrt{\sum_{i=0}^n a_i^2} \sqrt{\sum_{i=0}^n b_i^2} = -2a_0 b_0 + \sqrt{2a_0^2 - \beta} \sqrt{2b_0^2 - \beta}\\[-0.25in]
\]

\[
\quad\ \ \ =\frac{(2a_0^2-\beta)(2b_0^2-\beta)-4a_0^2b_0^2}{\sqrt{2a_0^2-\beta}\sqrt{2b_0^2-\beta}+2a_0b_0} =\frac{\frac{\beta^2}{a_0^2}-2\beta(1+\frac{b_0^2}{a_0^2})}{\sqrt{2-\frac{\beta}{a_0^2}}\sqrt{2\frac{b_0^2}{a_0^2}-\frac{\beta}{a_0^2}}+2\frac{b_0}{a_0}}\\[-0.07in]
\]
}
Since \(a_0, b_0 \gg \beta\) and \(a_0, b_0 \gg |a_0 - b_0|\), we have $\frac{\beta}{a_0^2}  \approx 0,\quad  \frac{b_0}{a_0}=\frac{a_0+(b_0-a_0)}{a_0}=1+\frac{b_0-a_0}{a_0} \approx 1$ and hence $\langle \mathbf{a}, \mathbf{b} \rangle \approx \frac{-2\beta \cdot 2}{2+2} = -\beta$.
Thus, the Lorentz distance is $\mathbf{d}_{Lo}(\mathbf{a}, \mathbf{b}) = |\langle \mathbf{a}, \mathbf{b} \rangle| - \beta \approx \beta - \beta = 0$.

\end{prf}
\vspace{-0.2in}
\end{theorem}

According to Theorem~\ref{lem:mapping1}, it becomes apparent that mapping two sets of points from Euclidean space to hyperbolic space, each set separated by the same Euclidean distance, results in distorted Lorentz distances. Furthermore, if the Euclidean distance of a pair of points in \(\mathcal{E}\) keeps constant and the pair  moves further from the origin, this angle diminishes, leading the Lorentz distance between the hyperbolic embeddings from the Vanilla Hyperbolic Projection to approach zero. Unfortunately, embeddings with similar directional properties typically arise from proximal objects, and hence identifying the most similar objects among those that are close remains a critical challenge in representation learning.

\vspace{-0.05in}
\subsection{Cosh Hyperbolic Projection}
\label{sec:projection_cosh}
\vspace{-0.03in}
\noindent \textbf{Motivation.} The limitation of the Vanilla Hyperbolic Projection stems from its mapping approach, which essentially translates the magnitude of the original Euclidean vector, $\sqrt{\sum_{i=1}^n x_i^2}$, into a hyperbolic coordinate, $x_{0} = \sqrt{\sum_{i=1}^n x_i^2 + \beta}$. This results in the mapped hyperbolic vectors becoming excessively large particularly when the magnitude is large, which tend to converge towards the hyperbolic asymptote line or the asymptotic hypersurface in higher dimensions as depicted by the gray dashed line in Figure~\ref{fig:mapping_err}. Consequently, the Lorentz distance between these vectors tends to be zero. To address this issue, a nonlinear transformation of the magnitudes is required to prevent excessive enlargement of hyperbolic space magnitudes relative to their Euclidean counterparts. Our proposed solution employs the hyperbolic cosine function, defined as $\cosh(d) = \frac{e^{d} + e^{-d}}{2}$, to moderate this expansion.

Actually, the hyperbolic cosine function \(\cosh(d)\) is derived from its definition via the hyperbolic angle. As illustrated in Figure~\ref{fig:cosh_mapping}, considering the orange point in the hyperbola \(x^2 - y^2 + 1 = 0\), its hyperbolic angle \(d\) is twice the area of the hyperbolic sector formed by the vector at this point and the y-axis (the orange-shaded area). The hyperbolic cosine and sine values, \(\cosh(d)\) and \(\sinh(d)\), of this hyperbolic angle correspond to the x and y coordinates of this point, respectively. (Note: the coordinates are reversed here for consistency with Figure~\ref{fig:mapping_err} to facilitate understanding). Utilizing the properties of the hyperbolic angle can significantly mitigate the issue of the Lorentz distance approaching zero due to the excessive magnitudes. To elucidate this further, we define the mapping process using the hyperbolic cosine function:
\vspace{-0.23in}

\begin{align*}
\small
\phi_{cosh}(\mathbf{x}^\mathcal{E}) \rightarrow \mathbf{x}^\mathcal{H}:\quad
&x_i^{\mathcal{H}} = 
\left\{
\begin{array}{ll}
\sqrt{\beta}\cosh(|x^{\mathcal{E}}|) & \text{if } i = 0, \\
x_{i-1}^{\mathcal{E}} \cdot k & others.
\end{array}
\right.
\\[-0.08in]
\label{equ:k}
k = \frac{\sinh(|x^{\mathcal{E}}|)}{|x^{\mathcal{E}}|}\cdot \sqrt{\beta} , &\qquad\  \ |x^{\mathcal{E}}| = \gamma_2\left(\sum_{i=0}^{n-1} (x_i^{\mathcal{E}})^2\right)\\[-0.34in]
\end{align*}

% \end{definition}

\noindent Here,\label{gamma} \(\gamma_c(x) = x^{\frac{1}{c}}\) represents a nonlinear compression of the Euclidean space distance. When \(c = 2\), we have \(\gamma_2(x) = \sqrt{x}\), and hence \(|x^\mathcal{E}|\) represents the magnitude.

\ifnum\tech=0 

\noindent \textbf{Effectiveness Analysis.} To demonstrate that the Cosh Hyperbolic Projection can address the distance degradation issue, we first prove in Theorem~\ref{lem:prf_cosh>vanilla} that the Lorentz distance in a 2-dimensional hyperbolic space remains non-diminished. 
\textcolor{black}{Then, we extend this conclusion to any $n+1$-dimensional space $(n>2)$ in Theorem~\ref{lem:not_zero} using Theorem~\ref{lem:dim_reduce}.}
% Then, in Theorem~\ref{lem:not_zero}, we extend this result to any \(n+1\)-dimensional space (\(n>2\)), effectively reducing it to the 2-dimensional case and thus completing the proof. 
% Notably, the proof of Theorem~\ref{lem:not_zero} relies on Lemma~\ref{lem:dim_reduce}, the proof of which is technical and therefore deferred to.

\else

\noindent \textbf{Effectiveness Analysis.} To demonstrate that the Cosh Hyperbolic Projection can address the distance degradation issue, we first prove in Theorem~\ref{lem:prf_cosh>vanilla} that the Lorentz distance in a 2-dimensional hyperbolic space remains non-diminished. 
\textcolor{black}{Then, in Theorem~\ref{lem:not_zero}, we extend this result to any \(n+1\)-dimensional space (\(n>2\)), effectively reducing it to the 2-dimensional case and thus completing the proof.}

\fi

\vspace{-0.06in}
\begin{theorem}\label{lem:prf_cosh>vanilla}
\(\forall \mathbf{a} = (a_0, a_1)\), \(\mathbf{b} = (b_0, b_1)\in\) \(\mathcal{H}_2(\beta)\), derived from the Cosh Hyperbolic Projection of the 1-dimensional values $a,b$ in the Euclidean space \(\mathcal{E}_{1}\), the Lorentz distance \(\mathbf{d}_{Lo}(\mathbf{a}, \mathbf{b})\) depends on \(|b - a|\) and non-diminished.
\vspace{0.01in}

\begin{prf}
\small
According to Definition of $\phi_{cosh}$, we have \(a_0 = \sqrt{\beta}\cosh(a), a_1 = \sqrt{\beta}\sinh(a)\), \(b_0 = \sqrt{\beta}\cosh(b)\) and \( b_1 = \sqrt{\beta}\sinh(b)\). Let \(b > a\) and \(d = b - a\), then we can conclude that:
\vspace{-0.18in}

\begin{align*}
\mathbf{d}_{Lo}(\mathbf{a}, \mathbf{b}) &= \left|\beta\cosh(a)\cosh(b) - \beta\sinh(a)\sinh(b)\right| - \beta \\[-0.07in]
&= \left|\beta\cosh(a - b)\right| - \beta = \beta(\frac{e^d + e^{-d}}{2} - 1)\\[-0.56in]
\end{align*}
\end{prf}
\end{theorem}

\vspace{-0.05in}
\begin{theorem}\label{lem:dim_reduce}
% \small
For any two vectors  \( \mathbf{a}, \mathbf{b}\in\mathcal{H}_{n+1}(\beta)\), if their components from the 1-st to the n-th dimension are linearly related, the $\mathbf{d}_{Lo}$ between them in \(\mathcal{H}_{n+1}(\beta)\) is consistent with that of two corresponding vectors in \(\mathcal{H}_2(\beta)\).

\ifnum\tech=0

% \else

\vspace{0.03in}
\begin{prf}
\small
For a vector  $\mathbf{a}=(a_0,a_1,\cdots,a_n) \in \mathcal{H}_{n+1}(\beta)= \{\mathbf{x} \in \mathbb{R}^{n+1} : \langle \mathbf{x}, \mathbf{x} \rangle = -\beta, x_0 > 1\}$, its corresponding vector that is linearly related to 1-st to n-th components can be represented as \(\mathbf{b} = (a_0 \frac{\sqrt{\sum_{i=1}^n a_i^2 \alpha^2 + \beta}}{\sqrt{\sum_{i=1}^n a_i^2 + \beta}}, \alpha a_1, ..., \alpha a_n)\). In addition, $\mathbf{a}$ and $\mathbf{b}$ can be considered as vectors $\mathbf{a}'=a_0\mathbf{e}_0+\sqrt{\sum_{i=1}^n a_i^2}\mathbf{e}_1$ and $\mathbf{b}'=a_0 \frac{\sqrt{\sum_{i=1}^n a_i^2 \alpha^2 + \beta}}{\sqrt{\sum_{i=1}^n a_i^2 + \beta}}\mathbf{e}_0+\alpha \sqrt{\sum_{i=1}^n a_i^2}\mathbf{e}_1$ in a 2-dimensional space spanned by the basis vectors \(\mathbf{e}_0\) and \(\mathbf{e}_1\), where \(\mathbf{e}_0 = (1, 0, 0, ..., 0)\) and \(\mathbf{e}_1 = \frac{(0, a_1, a_2, ..., a_n)}{\sqrt{\sum_{i=1}^n a_i^2}}\). In particular, $\mathbf{a}'$ and $\mathbf{b}'$ still lie on the 2-dimensional form of \(\mathcal{H}_2(\beta)= \{\mathbf{x} \in \mathbb{R}^2 : \langle \mathbf{x}, \mathbf{x} \rangle = -\beta\}\) according to the following equations:
\vspace{-0.1in}
{\footnotesize
\[
(a_0)^2- \left(\sqrt{\sum_{i=1}^n a_i^2}\right)^2=a_0^2-\sum_{i=1}^n a_i^2=\beta\\[-0.18in]
\]}

{\scriptsize\[
\hspace{-0.05in}\left(a_0 \frac{\sqrt{\sum_{i=1}^n a_i^2 \alpha^2 \hspace{-0.02in}+ \hspace{-0.02in} \beta}}{\sqrt{\sum_{i=1}^n a_i^2 \hspace{-0.02in}+ \hspace{-0.02in} \beta}}\right)^2\hspace{-0.05in}- \left( \alpha \sqrt{\sum_{i=1}^n a_i^2}\right)^2
\hspace{-0.1in}=\hspace{-0.02in}a_0^2\frac{\sum_{i=1}^n a_i^2 \alpha^2 \hspace{-0.02in}+ \hspace{-0.02in}\beta}{\sum_{i=1}^n a_i^2}-\alpha^2\sum_{i=1}^na_i^2\hspace{-0.02in}= \hspace{-0.02in}\beta.\\[-0.09in]
\]}
In particular, the distance $\mathbf{d}_{Lo}(\mathbf{a},\mathbf{b})$ in $\mathcal{H}_{n+1}(\beta)$ and $\mathbf{d}_{Lo}(\mathbf{a}',\mathbf{b}')$ in $\mathcal{H}_{2}(\beta)$ can be computed as follows:
\vspace{-0.07in}
{\small
\[
\mathbf{d}_{Lo}(\mathbf{a},\mathbf{b})=\big|a_0^2\frac{\sqrt{\sum_{i=1}^n a_i^2 \alpha^2 + \beta}}{\sqrt{\sum_{i=1}^n a_i^2 + \beta}}-\sum_{i=1}^n\alpha a_i^2\big| - \beta\\[-0.09in]
\]
\[
\mathbf{d}_{Lo}(\mathbf{a}',\mathbf{b}')=\big|a_0^2\frac{\sqrt{\sum_{i=1}^n a_i^2 \alpha^2 + \beta}}{\sqrt{\sum_{i=1}^n a_i^2 + \beta}}-\alpha \left(\sqrt{\sum_{i=1}^n a_i^2}\right)^2\big|-\beta\\[-0.11in]
\]
}

Therefore, we have $\mathbf{d}_{Lo}(\mathbf{a},\mathbf{b})=\mathbf{d}_{Lo}(\mathbf{a}',\mathbf{b}')$. Next, according to \textbf{Theorem \ref{lem:prf_cosh>vanilla}}, if $\mathbf{a}'$ and $\mathbf{b}'$ are mapped by $\phi_{cosh}$, we can deduce that $\mathbf{d}_{Lo}(\mathbf{a}',\mathbf{b}')$ is non-diminished, and hence the  $\mathbf{d}_{Lo}(\mathbf{a},\mathbf{b})$ is also non-diminished. 

\end{prf}
\vspace{-0.25in}
\fi

\end{theorem}

\begin{theorem}\label{lem:not_zero}
% \small
Given any two vectors in a hyperbolic space $\mathcal{H}_{n+1}(\beta)$ with the projection $\phi_{cosh}$ on  a Euclidean space $\mathcal{E}_{n}$, we can find a lower bound of the Lorentz distance between the two vectors, where the bound is non-dimished. 
% In using $\phi_{cosh}$ to map vectors in a Euclidean space $\mathcal{E}_{n}$ to the hyperbolic space $\mathcal{H}_{n+1}(\beta)$, for any 2 vector in n+1-dimensional $\mathcal{H}_{n+1}(\beta)$, $\mathbf{d}_{Lo}$ always has a none-zero lower boundary, avoiding $\mathbf{d}_{Lo}\rightarrow0$ caused by excessive norm.
\vspace{-0.02in}

\begin{prf}
\small
For Any vector $\mathbf{b}=(p_0,p_1,...p_n),\mathbf{c}=(q_0,q_1,...,q_n) \in \mathcal{H}_{n+1}(\beta)$,
We can always find a vector \(\mathbf{a} \in \mathcal{H}(\beta)\) that is linearly related to \(\mathbf{b}\) from the 1-st to \(n\)-th dimensions and identical to \(\mathbf{c}\) in the 0-th dimension as follows:
\vspace{-0.22in}
\[
\mathbf{a}=(q_0,kp_1,...,kp_n)\quad k = \sqrt{\frac{\sum_{i=1}^nq_i^2}{\sum_{i=1}^np_i^2}}\\[-0.08in]
\]
For the sake of simplicity, we can use \(\mathbf{a}\) to re-express \(\mathbf{b}\) and \(\mathbf{c}\) as follows:
\vspace{-0.21in}
{\footnotesize
\[
\mathbf{a} = (a_0,a_1,...,a_n),\  \mathbf{b}= (a_0\mu_{\mathbf{a}},\alpha a_1,...\alpha a_n), \  \mu_{\mathbf{a}}= \frac{\sqrt{\sum_{i=1}^na_i^2\alpha^2+\beta}}{\sqrt{\sum_{i=1}^na_i^2+\beta}}\\[-0.08in]
\]
}
{\footnotesize
\[
\mathbf{c}= \left(a_0,a_1(1+\delta_1),...,a_{n-1}(1+\delta_{n-1}),\sqrt{a_n^2 -\sum_{i=1}^{n-1}(\delta_i^2+2\delta_i)a_i^2}\right)\\[-0.08in]
\]
}
where we use \(p_i\) and \(q_i\) to inversely solve for the relevant parameters:
\vspace{-0.25in}

\[
\alpha =\frac{1}{k}, \quad \delta_i=\frac{q_i}{kp_i}-1,\quad a_0=q_0, \quad a_i=kp_i\\[-0.08in]
\]
Next, we can compute \(\mathbf{d}_{Lo}(\mathbf{b},\mathbf{c})\) for the arbitrary vectors $\mathbf{b}, \mathbf{c}$ and  \(\mathbf{d}_{Lo}(\mathbf{b},\mathbf{a})\) for $\mathbf{a}, \mathbf{b}$ as follows:
\vspace{-0.16in}
{\scriptsize
\[
\mathbf{d}_{Lo}(\mathbf{a},\mathbf{b})=|a_0^2\mu_{\mathbf{a}}-\alpha\sum_{i=1}^na_i^2|-\beta \\[-0.1in]
\]
}
{\scriptsize
\[
\mathbf{d}_{Lo}(\mathbf{b},\mathbf{c})=\left|a_0^2\mu_{\mathbf{a}}-\sum_{i=1}^{n-1}\alpha(1+\delta_i)a_i^2-\alpha a_n^2\sqrt{1-\sum_{i=1}^{n-1}(\delta_i^2+2\delta_i)\frac{a_i^2}{a_n^2}}\right|-\beta\\[-0.08in]
\]}
Thus, we can demonstrate that \(\mathbf{d}_{Lo(\mathbf{b},\mathbf{c})} \geq \mathbf{d}_{Lo(\mathbf{b},\mathbf{a})}\) as follows:
\vspace{-0.08in}
{\scriptsize
\[
\mathbf{d}_{Lo}(\mathbf{b},\mathbf{c})-\mathbf{d}_{Lo}(\mathbf{b},\mathbf{a})=\alpha a_n^2\left(1-\sqrt{1-\sum_{i=1}^{n-1}\frac{a_i^2}{a_n^2}(\delta_i^2+2\delta_i)}-\sum_{i=1}^{n-1}\frac{a_i^2}{a_n^2}\delta_i\right)\\[-0.08in]
\]
\[
\qquad\qquad\qquad\quad\ \ \,\geq\alpha a_n^2\left(1-\sqrt{1-2\sum_{i=1}^{n-1}\frac{a_i^2}{a_n^2}\delta_i}-\sum_{i=1}^{n-1}\frac{a_i^2}{a_n^2}\delta_i\right)\geq 0 \\[-0.08in]
\]
}
In addition, according to \textbf{Lemma \ref{lem:dim_reduce}}, $\mathbf{d}_{Lo}(\mathbf{b},\mathbf{a})$ cannot collapse to zero due to they are linearly related from the $1$-th to the $n$-th dimension. Thus, if using $\phi_{cosh}$, for any vector $\mathbf{b},\mathbf{c}$ in $\mathcal{H}$, their $\mathbf{d}_{Lo}(\mathbf{b},\mathbf{c})$ has its lower boundary and non-diminished.

\end{prf}
\vspace{-0.25in}
\end{theorem}

The advantage of this approach lies in the proportionality between the input to the hyperbolic cosine function and the area of the shaded region in Figure~\ref{fig:cosh_mapping}. Specifically, the red Euclidean distance \(d\) and the hyperbolic sector (blue shaded area) are linearly correlated. As the Euclidean magnitude increases, the area of the hyperbolic sector also increases proportionally, causing the red angle in the figure to increase correspondingly. This proportional increase significantly mitigates the issue of the Lorentz distance approaching zero due to large magnitudes.

\noindent \textbf{Complexity Analysis of Hyperbolic Projection.} The complexity of our projection layer is linear to the dimensionality $d$ of the vectors, denoted as \(O(d)\). This time complexity is negligible compared to existing embedding models. For instance, the embedding complexities of Neutraj and TrajGAT are \(O(Ld^2)\), which is significantly greater than \(O(d)\), where \(L\) is the average number of trajectory points.

\vspace{-0.06in}
\section{Dynamic Fusion Distance}
\label{sec:Fusion_dist}
\vspace{-0.04in}
In this section, we first introduce the variability of triangle constraints in different trajectory similarity functions in Section~\ref{subsec:triag_vio}. To enhance the robustness of trajectory embeddings by addressing this variability, we design a dynamic fusion distance module that integrates Euclidean distance and Lorentz 

\noindent distance in Section~\ref{subsec:dfd}.

\vspace{-0.07in}
\subsection{Triangle Constraint Violation Variability}
\label{subsec:triag_vio}
\vspace{-0.04in}

Given a specific trajectory dataset \(\mathcal{T}\), the distances measured by different similarity functions, such as DTW and EDR, tolerate varying degrees of inequality constraint violations. To quantitatively analyze this variability, we use a violation flag to label whether a tuple distance follows the inequality constraint as follows:
\vspace{-0.22in}

\begin{equation*}
\textbf{TVF}\scriptsize(T_i, T_j, T_k) = 
\begin{cases} 
  1 & \text{if} \max(Sim[k | i, j], Sim[i | j, k], Sim[j | i, k]) > 0 \\
  0 & \text{otherwise}
\end{cases}
\vspace{-0.3in}
\end{equation*}

\begin{equation*}
Sim[k | i, j] = f(T_i, T_j) - f(T_i, T_k) - f(T_j, T_k)\quad \\[-0.05in]
\end{equation*}

\noindent where \(f(T_i, T_j)\) denotes a given trajectory similarity function.

Next, we propose two metrics, namely the \textit{Ratio of Violation} and \textit{Average Relative Violation}, to represent the ratio of violating constraints and the average degree of violation, respectively.

\vspace{-0.05in}
\begin{definition}[Ratio of Violation]
The ratio of violation for a given dataset \(\mathcal{T}\) is denoted as \(\textbf{RV}(\mathcal{T})\), and is computed as follows:
\vspace{-0.3in}

\begin{equation*}
\small
  \textbf{RV}(\mathcal{T}) = \sum_{T_i, T_j, T_k \in \mathcal{T}} \frac{\textbf{TVF}(T_i, T_j, T_k)}{\mathcal{C}_N^3} \quad \quad (N = |\mathcal{T}|)\\[-0.1in]
\end{equation*}
\end{definition}

\begin{definition}[Average Relative Violation]
The average relative violation is denoted as \(\textbf{ARVS}(\mathcal{T})\), and is computed as follows:
\vspace{-0.27in}

\small
\begin{align*} 
\small
&\textbf{ARVS}(\mathcal{T})=\frac{\sum_{T_i,T_j,T_k\in\mathcal{T}} \textbf{RVS}(T_i,T_j,T_k)\textbf{TVF}(T_i,T_j,T_k)}{\sum_{T_i,T_j,T_k\in\mathcal{T}}\textbf{TVF}(T_i,T_j,T_k)}
\\[-0.3in]
\end{align*}
\begin{align*} 
\vspace{0.03in}
&\textbf{RVS}\scriptsize{(T_i,T_j,T_k) =} 
    \begin{cases} 
        \frac{\displaystyle Sim[ k | i, j ]}{\displaystyle f(T_k, T_i)+f(T_k, T_j)} & \text{if $f(T_i, T_j)$ is max.} \\
        \frac{\displaystyle Sim[ i | j, k ]}{\displaystyle f(T_i, T_j)+f(T_i, T_k)} & \text{if $f(T_j, T_k)$ is max.} \\
        \frac{\displaystyle Sim[ j | i, k ]}{\displaystyle f(T_j, T_k)+f(T_j, T_i)} & \text{if $f(T_i, T_k)$ is max.}
    \end{cases}
\end{align*}
\vspace{-0.26in}
\end{definition}

% \marginpar{\vspace{-1.3cm}\scriptsize{\textcolor{blue}{R3.D8}}}
\color{black}
\begin{example}
% \marginpar{\vspace{0.2cm}\scriptsize{\textcolor{blue}{R3.D2}}}
\label{exmp3}
Consider a trajectory dataset containing four trajectories $\mathcal{T}=\{T_a,T_b,T_c,T_d\}$, where only the similarity distances between $T_a, T_b, T_c$ do not satisfy the triangle inequality, with $f(T_a,T_b)=5$, $f(T_a,T_b)=2$, and $f(T_a,T_b)=1$. Therefore, $sim[c|a,b]=2$, and $\textbf{TVF}(T_a,T_b,T_c)=1$, where $\textbf{TVF}$ indicates that this trajectory triplet does not satisfy the triangle inequality constraint. Then, $\textbf{RV}(\mathcal{T})=\frac{1}{4}$, which means that $\mathcal{T}$ has one-fourth of its samples violating the triangle constraint. Finally, $\textbf{ARVS}(\mathcal{T})=\frac{2}{3}$, indicating that among the samples violating the triangle constraint, the average relative violation size is $\frac{2}{3}$. The larger this relative violation, the worse the performance of the Euclidean distance on it.
\end{example}
\vspace{-0.02in}
\color{black}

\textcolor{black}{As reported in Table~\ref{tbl:constraint_dataset}, we computed the three metrics on the commonly used datasets~\cite{chengdu, porto, tdrive, osm, geolife}. The results demonstrate the variability of different similarity functions across these datasets. According to the definition of \textbf{RV}, approximately above one-fifth of the trajectory triplets on the DTW metric violate the triangle inequality constraint. It is a proportion that cannot be ignored.}
% \marginpar{\vspace{-2.7cm}\scriptsize{\textcolor{blue}{R3.D4}}}
Additionally, these large \textbf{ARVS} indicates the challenge of accurately representing true similarity using embeddings in Euclidean space. To address this, leveraging hyperbolic space provides a more effective representation of these trajectories.

% \begin{table}[!t]
%     \centering
%     \caption{Constraint Variability on Commonly Trajectory datasets}
%     \setlength{\tabcolsep}{3pt}
%     \label{tbl:constraint_dataset}

%     \begin{adjustbox}{center} 
%     \resizebox{1\linewidth}{!}{
%     % \scalebox{0.87}{
%     \begin{tabular}{c|ccc|ccc}
%     \toprule
%     \toprule
%     \multicolumn{1}{c|}{\textbf{Dataset}} & \multicolumn{3}{c|}{\textbf{Chengdu}} & \multicolumn{3}{c}{\textbf{Porto}} \\ \hline

%     Method & DTW & SSPD & EDR & DTW & SSPD & EDR \\ \hline
%      \textbf{RV} & 19.26\% & 28.61\% & 13.02\% & 25.28\% & 27.84\% & 16.66\%\\ \hline
%     \textbf{ARVS} & 0.1469 & 0.1252 & 0.2327 & 0.1590 & 0.1211 & 0.3179 \\ \bottomrule
%     \bottomrule
%     \end{tabular}
%     }
%     \end{adjustbox}
%     \vspace{-0.15in}
% \end{table}

\begin{table}[!t]
\vspace{-0.1in}
    \centering
    \caption{\textcolor{black}{Constraint Variability on Trajectory datasets}}
    \vspace{-0.09in}
    \setlength{\tabcolsep}{3pt}
    \label{tbl:constraint_dataset}

    \begin{adjustbox}{center} 
    \resizebox{0.95\linewidth}{!}{
    % \scalebox{0.87}{
    \begin{tabular}{c|c|cccccc}
    \toprule
    \toprule
    \multicolumn{2}{c}{\textbf{Dataset}} & \textbf{Chengdu} & \textbf{Porto} & \textbf{Xian}& \textbf{T-Drive}& \textbf{OSM}& \textbf{Geolife} \\ \hline

    \multicolumn{2}{c|}{Trajectory number} & 5.2m & 1.5m & 7.3m & 113.4k & 11.7m & 17.5k \\ \hline
    \multirow{2}{*}{DTW} & \textbf{RV} & 19.3\% & 25.3\% & 20.7\% & 36.9\% & 15.4\% & 38.0\%\\ \cline{2-8} 
    &\textbf{ARVS} & 0.147 & 0.159 & 0.103 & 0.486 & 0.041 & 0.144 \\ \hline
    
    \multirow{2}{*}{SSPD} & \textbf{RV} & 28.6\% & 27.8\% & 22.6\% & 37.0\% & 5.7\% & 18.6\%\\ \cline{2-8} 
    &\textbf{ARVS} & 0.125 & 0.121 & 0.057 & 0.126 & 0.048 & 0.044 \\ \hline
    
    \multirow{2}{*}{EDR} & \textbf{RV} & 13.0\% & 16.7\% & 38.2\% & 53.7\% & 9.4\% & 11.8\%\\ \cline{2-8} 
    &\textbf{ARVS} & 0.233 & 0.318 & 1.087 & 1.427 & 0.166 & 1.756 \\ \bottomrule        \bottomrule
    \end{tabular}
    }
    \end{adjustbox}
    \vspace{-0.25in}
\end{table}

% \begin{table}[!t]
%     \centering
%     \caption{Constraint Variability on Chengdu and Porto}
%     \setlength{\tabcolsep}{3pt}
%     \label{tbl:constraint_dataset}

%     \begin{adjustbox}{center} 
%     \resizebox{1\linewidth}{!}{
%     % \scalebox{0.87}{
%     \begin{tabular}{c|ccc|ccc}
%     \toprule
%     \toprule
%     \multicolumn{1}{c|}{\textbf{Dataset}} & \multicolumn{3}{c|}{\textbf{Chengdu}} & \multicolumn{3}{c}{\textbf{Porto}} \\ \hline

%     Method & DTW & SSPD & EDR & DTW & SSPD & EDR \\ \hline
%      \textbf{RV} & 19.26\% & 28.61\% & 13.02\% & 25.28\% & 27.84\% & 16.66\%\\ \hline
%     \textbf{ARVS} & 0.1469 & 0.1252 & 0.2327 & 0.1590 & 0.1211 & 0.3179 \\ \bottomrule
%     \bottomrule
%     \end{tabular}
%     }
%     \end{adjustbox}
%     \vspace{-0.15in}
% \end{table}

\vspace{-0.08in}
\subsection{Distance Fusion}
\label{subsec:dfd}
\vspace{-0.05in}
\noindent\textbf{Motivation.} 
% \marginpar{\vspace{0.1cm}\scriptsize{\textcolor{blue}{R1.O3}}}
Although some cases break the constraints, many

\noindent  data points still adhere to the inequality constraint. 
% \textcolor{blue}{Due to the limited precision of float numbers, the values of the Hyperbolic space and Lorentz distance can only be approximated}. 
Therefore, relying solely on the Lorentz distance in the mapped hyperbolic space is ineffective; the Euclidean distance should also be considered. Consequently, given the variability of the constraints, we advocate for a dynamic method to identify trajectories constrained by the triangle inequality. This approach allows for the adaptive allocation of Lorentzian and Euclidean distances based on the varying magnitudes of the constraints.
% \marginpar{\vspace{-6cm}\scriptsize{\textcolor{blue}{R3.D4\\R3.D5}}}

To design an effective dynamic fusion procedure, it is crucial to avoid introducing additional computational complexity. For instance, a model that generates a weighting factor for the Lorentzian distance relative to the Euclidean distance for each pair of input trajectories would need to process every pair, resulting in quadratic time complexity. This would drastically increase the time required for similarity computation, contradicting the goal of enhancing speed using trajectory representation learning.

Instead, we propose a method that maintains the linear time complexity of the original pipeline. We achieve this by generating a Lorentz factor vector and a Euclidean factor vector for each trajectory. The proportion of the Lorentzian distance between any two trajectories is then determined by the ratio of the inner product of their Lorentz factor vectors to the inner product of their Euclidean factor vectors as follows:
\vspace{-0.37in}

\begin{equation*}
\label{equ:alpha}
\begin{aligned}
% \label{equ:dfm1}
V^{Lo}_a, V^{Eu}_a & = \textbf{Encoder}(T_a) \qquad
V^{Lo}_b, V^{Eu}_b = \textbf{Encoder}(T_b) \\[-0.04in]
\alpha^{Lo}_{a,b} &= \frac{V^{Lo}_a \cdot V^{Lo}_b}{V^{Lo}_a \cdot V^{Lo}_b + V^{Eu}_a \cdot V^{Eu}_b}\\[-0.06in]
\end{aligned}
\end{equation*}

\noindent where the \textbf{Encoder} is a sequence-to-vector model that accepts trajectory sequences and generates a vector. The first half of the vector, denoted as \(V^{Lo}_*\), represents the Lorentz factor embedding, while the second half, denoted as \(V^{Eu}_*\), represents the Euclidean factor embedding. The inner product of these vectors between different trajectories produces the Lorentz proportion \(\alpha^{Lo}\), which balances the weights of the Lorentzian and Euclidean distances. The process of generating a dynamic fused distance from trajectory embeddings is as follows:
\vspace{-0.21in}

\begin{equation*}
\small
\label{equ:dfm4}
\textbf{d}_{Fu}(T_a, T_b) = \alpha^{Lo}_{a,b} \cdot \textbf{d}_{Lo}(\textbf{x}_a^{\mathcal{H}}, \textbf{x}_b^{\mathcal{H}}) + (1 - \alpha^{Lo}_{a,b}) \cdot \textbf{d}_{Eu}(\textbf{x}_a^{\mathcal{E}}, \textbf{x}_b^{\mathcal{E}})
\end{equation*}

\vspace{-0.14in}
\color{black}
\begin{example}
% \marginpar{\vspace{0.1cm}\scriptsize{\textcolor{blue}{R1.O3}}}
\label{exmp2}
There are two sets of trajectories, $\{T_a,T_b,T_c\}$ and $\{T_d,T_e,T_f\}$. The former severely violates the triangle inequality constraint, $\textbf{RVS}(T_a,T_b,T_c)>1$; while the latter satisfies such constraint, $\textbf{TVF}(T_i, T_j, T_k) = 0$. 
When calculating the similarity distance $\textbf{d}_{Fu}$ between $T_a$ and $T_b$ (or $T_c$), a larger proportion of Lorentz distance should be used, while for the $\textbf{d}_{Fu}$ calculation between $\{T_d, T_e, T_f\}$, a larger proportion of Euclidean distance should be adopted. 

These trajectories will be processed through the factor embeddings (\(V^{Lo}_*, V^{Eu}_*\)) generated by the \textbf{Encoder($\cdot$)} to compute their Lorentz proportions (\(\alpha^{Lo}\)). The \textbf{Encoder($\cdot$)} will learn the triangle inequality constraint situations between different trajectories and produce appropriate factor embeddings \(V^{Lo}_*, V^{Eu}_*\), so that \(\alpha^{Lo}_{a,b} >> \alpha^{Lo}_{d,e}\). Ultimately, this process achieves the dynamic allocation of the Lorentz distance ratio adapting to the triangle constraint.
\end{example}
\color{black}
\vspace{-0.07in}

As for the model selection of \textbf{Encoder}, we experimented with several common Seq2Vec models. To accommodate various trajectory embedding models without introducing additional time complexity, we selected the LSTM, which has the lowest complexity and scales linearly with trajectory length. Since this model is straightforward and not the primary focus of this paper, we will not elaborate further.

\vspace{-0.01in}
\noindent \textbf{Complexity Analysis.}  The Lorentz distance \(\textbf{d}_{Lo}\), similar to the Euclidean distance \(\textbf{d}_{Eu}\), is calculated as the dot product between vectors, with a computational complexity of \(O(d)\), where \(d\) is the dimensionality. Therefore, the overall complexity of computing the fused distance $\textbf{d}_{Fu}$ remains \(O(d)\).

\begin{table}[!t]
\vspace{-0.1in}
\caption{Different Embedding Models and Network Types}
\label{tbl:net}
\vspace{-0.09in}
\setlength{\tabcolsep}{5.5pt}
\centering
\begin{adjustbox}{center} 
\resizebox{0.91\linewidth}{!}{
% \scalebox{0.90}{
\begin{tabular}{c|c|c|c|c}
\toprule
\toprule
Embedding Model & RNN & Attn & GNN & Preprocess \\\hline
TrajGAT &  & \checkmark & GAT & Quad-Tree \\ \hline
Neutraj & GRU &  & & Grid-Cell  \\ \hline
Traj2SimVec & LSTM &  & & Sub-Trajectory\\ \hline
ST2Vec & LSTM & \checkmark & & \\ \bottomrule
\bottomrule
\end{tabular}}
\end{adjustbox}
\vspace{-0.25in}
\end{table}

\vspace{-0.02in}
\section{Experiments}\label{sec:exp}
\vspace{-0.05in}

\iffalse
In this chapter, we evaluated the effectiveness and efficiency of several mainstream state-of-the-art trajectory similarity models after incorporating the \textbf{LH}-plugin across several commonly used public real-world datasets. 
Additionally, we analyzed the impact of data scale and training stability. Finally, we conducted ablation experiments on various modules within the \textbf{LH}-plugin, thereby substantiating the efficacy of the \textbf{LH}-plugin and our three principal contributions. 

To ensure that performance improvements are solely attributable to the \textbf{LH}-plugin, all hyper-parameters of the original models remain unchanged, using the default settings as specified in their open-source code repositories. 
\fi

In this section, we conduct extensive experiments to evaluate our proposed framework \textbf{LH}-plugin. In particular, we focus on answering the following questions.
\vspace{-0.01in}
\begin{itemize}[leftmargin=12pt]
    \item \textbf{Q1}: Can the \textbf{LH}-plugin address the triangle inequality issue in trajectory embeddings?
    \textbf{Q1.1 (Section~\ref{q1.1})}: Do embeddings generated with \textbf{LH}-plugin significantly enhance accuracy in trajectory similarity retrieval?
    \textbf{Q1.2 (Section~\ref{q1.2})}: Does \textbf{LH}-plugin bring the triangular relationships between embeddings closer to the ground truth, thereby relaxing the triangle inequality constraint?
    
    \item \textbf{Q2}: How does the \textbf{LH}-plugin influence the training and inference efficiency of embedding models? \textbf{Q2.1 (Section~\ref{q2.1})}: Does the increased inference latency affect the efficiency of similarity trajectory retrieval? \textbf{Q2.2 (Section~\ref{q2.2})}: What is the scalability of the \textbf{LH}-plugin framework for training embedding models? 
    
    \item \textbf{Q3 (Section~\ref{q2.3})}: Is the \textbf{LH}-plugin framework robust in enhancing similarity retrieval effectiveness across varying model effectiveness?
    
    \item \textbf{Q4 (Section~\ref{q3})}: Quantitative analysis of our three principal contributions in the \textbf{LH}-plugin.

    \item  \textbf{Q5 (Section~\ref{q5})}: Quantitative analysis of hyper-parameters.
\end{itemize}

\vspace{-0.1in}
\subsection{Experimental Setup}
\vspace{-0.05in}
\noindent\textbf{Datasets.} We select several datasets commonly used by trajectory similarity methods for experiments.

(1) \textbf{Chengdu}~\cite{chengdu}: The Chengdu dataset comprises 5.22 million ride-hailing trajectories, situated in Chengdu, China, provided by DiDi Inc. It includes a total of 1.05 billion coordinate points, with an average trajectory length of 32.5 km, which is commonly used for spatial similarity measures.

(2) \textbf{Porto}~\cite{porto}: The Porto dataset consists of 1.49 million taxi route trajectories, located in Porto, Portugal, and is open source on Kaggle. It encompasses over 80 million points, with an average trajectory length of 6.12 km, which is commonly used for spatial similarity measures.

(3) \textbf{T-Drive}~\cite{tdrive}: The T-Drive is a spatio-temporal trajectory dataset that includes trajectories from 10,357 taxis over one week. It consists of approximately 15 million data points, covering a total distance of 9 million kilometers.

\begin{table*}[t]
\vspace{-0.06in}
\centering
\caption{Accuracy results on the Chengdu and Porto datasets}
\vspace{-0.09in}
\label{tbl:acc}
\setlength{\tabcolsep}{3pt}
\begin{adjustbox}{center} 
\resizebox{1\textwidth}{!}{
\begin{tabular}{c|c|lccccc|c|lccccc}
\toprule
\toprule
\multirow{2}{*}{base model} & \multicolumn{7}{c|}{Chengdu} & \multicolumn{7}{c}{Porto} 
\\
\cline{2-15} 
& sim & plugin & HR@5(\%) & HR@10(\%) & HR@50(\%) & NDCG10 & NDCG50 & sim & plugin &  HR@5(\%) & HR@10(\%) & HR@50(\%) & NDCG10 & NDCG50 \\
\hline
\multirow{9}{*}{Neutraj} & \multirow{3}{*}{DTW} & Original & 53.54 & 60.36 & 75.28 & 0.7526 & 0.8513&\multirow{3}{*}{DTW} & Original & 36.8 & 41.82 & 52.09 & 0.4487& 0.5452 \\
& & \textbf{LH}-plugin & 68.33 & 73.99 & 84.11 & 0.7929 & 0.8790 & & \textbf{LH}-plugin & 39.79 & 44.31 & 54.41 & 0.4916 & 0.5978 \\
& & \%Increase & +27.62\% & +22.58\% & +11.73\% & +5.35\% & +3.25\% & & \%Increase & +8.13\% & +5.95\% & +4.45\% & +9.56\% & +9.65\% \\
\cline{2-15} 
& \multirow{3}{*}{SSPD} & Original & 49.80 & 54.48 & 66.81 & 0.6112 & 0.7293 & \multirow{3}{*}{SSPD} & Original & 23.61 & 27.12 & 38.14 & 0.3046 & 0.4125\\
& & \textbf{LH}-plugin & 57.42 & 63.51 & 78.52 & 0.6484 & 0.8028 & & \textbf{LH}-plugin & 27.26 & 31.30 & 43.27 & 0.3611 & 0.4943\\
& & \%Increase & +15.30\% & +16.57\% & +17.53\% & +6.09\% &+10.08\% & & \%Increase & +15.46\% & +15.41\% & +13.45\% & +18.55\% & +19.83\% \\

\cline{2-15} 
& \multirow{3}{*}{EDR} & Original & 8.81 & 10.71 & 13.21 & 0.1373 &0.1730 & \multirow{3}{*}{EDR} & Original & 9.54 & 10.23 & 11.60 & 0.1103 & 0.1268 \\
& & \textbf{LH}-plugin & 12.06 & 14.46 & 1.48 & 0.1546 & 0.1968& & \textbf{LH}-plugin & 11.48 & 12.18 & 13.83 & 0.1539 & 0.1579 \\
& & \%Increase & +36.89\% & +35.01\% & +32.32\% & +12.60\% & +13.76\% & & \%Increase & +20.34\% & +19.06\% & +19.22\% & +39.53\% & +24.53\% \\

\hline
\hline
\multirow{9}{*}{TrajGAT} & \multirow{3}{*}{DTW} & Original & 27.07 & 30.31 & 36.69 & 0.3526 & 0.4843 & \multirow{3}{*}{DTW} & Original & 14.42 & 18.39 & 33.71 & 0.1422 & 0.2775 \\
& & \textbf{LH}-plugin & 39.92 & 41.95 & 47.32 & 0.4877 & 0.5969 & & \textbf{LH}-plugin & 17.24 & 21.52 & 35.66 & 0.1898 & 0.3605 \\
& & \%Increase & +47.47\% & +38.40\% & +28.97\% & +38.32\% &+23.25\% & & \%Increase & +19.56\% & +17.02\% & +5.78\% & +33.47\% & +29.91\% \\

\cline{2-15} 
& \multirow{3}{*}{SSPD} & Original & 38.61 & 43.73 & 55.37 & 0.4753 & 0.6033 & \multirow{3}{*}{SSPD} & Original & 20.07 & 25.35 & 37.15 & 0.2458 & 0.3515 \\
& & \textbf{LH}-plugin & 45.1 & 50.33 & 61.74 & 0.5455 & 0.6832 & & \textbf{LH}-plugin & 20.5 & 25.41 & 40.77 & 0.2562 & 0.3641 \\
& & \%Increase & +16.81\% & +15.09\% & +11.50\% & +14.77\% & +13.24\% & & \%Increase & +2.14\% & +0.24\% & +9.83\% & +4.23\% & +3.58\% \\

\cline{2-15} 
& \multirow{3}{*}{EDR} & Original & 5.74 & 6.63 & 8.19 & 0.0473 & 0.0737 & \multirow{3}{*}{EDR} & Original & 7.42 & 8.03 & 9.78 & 0.0775 & 0.0861 \\
& & \textbf{LH}-plugin & 6.69 & 7.49 & 8.65 & 0.0725 & 0.0817 & & \textbf{LH}-plugin & 7.46 & 8.10 & 9.81 & 0.0783 & 0.0892\\
& & \%Increase & +16.55\% & +12.97\% & +5.62\% & +53.28\% & +10.85\% & & \%Increase & +0.54\% & +0.87\% & +0.31\% & +1.03\% & +3.60\% \\

\hline
\hline
\multirow{9}{*}{Traj2SimVec} & \multirow{3}{*}{DTW} & Original & 44.87 & 50.38 & 65.41 & 0.5071 & 0.6559 & \multirow{3}{*}{DTW} & Original & 19.94 & 24.31 & 37.45 & 0.2771 &0.4226 \\
& & \textbf{LH}-plugin & 47.11 & 53.3 & 68.4 & 0.5545 & 0.6990 &  & \textbf{LH}-plugin & 28.91 & 33.68 & 49.91 & 0.3239 & 0.4753 \\
& & \%Increase & +4.99\% & +5.80\% & +4.57\% & +9.35\% & +6.57\% & & \%Increase & +44.98\% & +38.54\% & +33.27\% & +16.89\% &+12.47\%\\

\cline{2-15} 
& \multirow{3}{*}{SSPD} & Original & 47.68 & 52.64 & 66.05 & 0.6092 & 0.7348 & \multirow{3}{*}{SSPD} & Original & 32.00 & 36.85 & 51.25 & 0.4374 &0.5893 \\
& & \textbf{LH}-plugin & 57.37 & 62.99 & 74.04 & 0.6818 & 0.7856 & & \textbf{LH}-plugin & 38.29 & 44.81 & 60.86 & 0.4694 & 0.6139 \\
& & \%Increase & +20.32\% & +19.66\% & +12.10\% & +11.92\% & +6.91\% & & \%Increase & +19.66\% & +21.60\% & +18.75\% & +7.32\% & +4.17\% \\

\cline{2-15} 
& \multirow{3}{*}{EDR} & Original & 11.9 & 14.51 & 17.56 & 0.1746 & 0.2126 &\multirow{3}{*}{EDR} & Original & 20.33 & 20.52 & 17.54 & 0.2116 & 0.2083 \\
& & \textbf{LH}-plugin & 15.89 & 19.68 & 25.08 & 0.1979 & 0.2312 & & \textbf{LH}-plugin & 20.71 & 21.36 & 19.00 & 0.2230 & 0.2133 \\
& & \%Increase & +33.53\% & +35.63\% & +42.82\% & +13.34\% & +8.75\% & & \%Increase & +1.87\% & +4.09\% & +8.32\% & +5.39\% &+2.40\%\\

\bottomrule
\bottomrule
\end{tabular}
}
\end{adjustbox} 
\vspace{-0.19in}
\end{table*}

\noindent\textbf{Embedding models.} To evaluate the adaptability of the \textbf{LH}-plugin across various scenarios, we select four advanced trajectory embedding models  as follows:
% We evaluated the performance of the \textbf{LH}-plugin using the following four methods:
\vspace{-0.04in}

\begin{itemize}[leftmargin=10pt]
  \item \textbf{Neutraj}~\cite{neutraj} is a spatial trajectory embedding model that utilizes LSTM networks and a gridded neighbor table to store and update proximity information between trajectory points.
  \item \textbf{TrajGAT}~\cite{TrajGAT} is designed for handling spatial similarity computation in long trajectories. In particular, it employs a pre-constructed quadtree topology and integrates graph attention networks to save GPU memory and enhance performance for long trajectories. 
  \item \textbf{Traj2SimVec}~\cite{traj2s} is another representation learning model for spatial trajectory similarity. It introduces innovative sub-trajectory and trajectory matching information, achieving high performance in pure spatial tasks.
  \item \textbf{ST2Vec}~\cite{st2vec} addresses similarity issues in spatio-temporal trajectories by incorporating time encoding into trajectory encoding and combining them through co-attention mechanisms.
  \item \textcolor{black}{\textbf{Tedj}~\cite{tedj} addresses the issues of GPS sampling rate fluctuations and sampling point offsets in spatio-temporal trajectories by using a 3D spatio-temporal grid method to transform the trajectories into spatio-temporal grid-related sequences. }
  % \marginpar{\vspace{-2cm}\scriptsize{\textcolor{blue}{R2.O2}}}
\end{itemize}
\vspace{-\topsep}

In summary, the select models represent the most commonly employed neural networks for trajectory embedding, including RNN variants, attention mechanisms, and graph neural networks. Additionally, these models utilize a diverse array of trajectory preprocessing methods, such as sub-trajectory segmentation, grid-cell partitioning, and tree-based indexing structures as shown in Table~\ref{tbl:net}.

% \noindent \textit{Remark.} 
% \textcolor{red}{In order to evaluate the adaptability of the \textbf{LH}-plugin to different neural network models and to various trajectory preprocessing approaches, we carefully selected the four aforementioned baseline models from a large set of trajectory learning-based similarity methods. These baseline models encompass the most commonly used neural networks in trajectory embedding (RNN-Variants, Attention mechanisms, and Graph Neural Networks) while also covering a wide range of widely utilized trajectory preprocessing methods, including sub-trajectory segmentation, grid-cell partitioning, and tree-based indexing structures as shown in Table~\ref{tbl:net}.}

% For spatial trajectory embedding models(i.e., \textbf{Neutraj}, \textbf{TrajGAT}, \textbf{Traj2SimVec}), we used the Chengdu and Porto datasets, and standardized the preprocessing by removing duplicate trajectory points. For the spatio-temporal trajectory embedding model \textbf{ST2Vec}, we employed T-Drive to compare the performance in spatio-temporal trajectory similarity, validating the feasibility of \textbf{LH}-plugin in spatio-temporal trajectories.
\smallskip

\vspace{-0.05in}
\noindent\textbf{Evaluation Metrics.} For the spatial similarity computation, we select \textbf{DTW}, \textbf{SSPD}, and \textbf{EDR} as benchmark similarity measures, which do not satisfy the triangle inequality constraint. For spatio-temporal trajectory similarity, we employ \textbf{TP}, \textbf{DITA}, and \textbf{secret Fréchet} distance to assess the effectiveness of the \textbf{LH}-plugin. 

\vspace{-0.01in}
To gauge accuracy, we follow established practices~\cite{neutraj, TrajGAT, st2vec}, using hit rate $HR@\alpha$ 
and NDCG~\cite{ndcg} 
as evaluation metrics, where higher values indicate more precise trajectory similarity computation. Here, HR@$\alpha$ represents the hit rate of top $\alpha$, indicating the proportion of the most similar top $\alpha$ in the embedded results relative to the most similar top $\alpha$ in the ground truth. 
While, NDCG~\cite{ndcg} reflects the similarity between the ranking of the embedded results and the ranking of the\, ground\, truth,\, and\, is\, commonly\, used\, to\, measure\, the 

\noindent retrieval quality in search and recommendation systems.

\noindent\textbf{Environment Settings.} We implement and incorporat the \textbf{LH}-plugin based on PyTorch 2.0.1 into these original trajectory similarity models. Additionally, the experiments are conducted on Ubuntu 18.04.6 OS, utilizing the Intel Xeon Gold 6248 CPU and the NVIDIA Tesla V100 GPU, yielding the following experimental results.

\vspace{-0.03in}
\subsection{Effectiveness Evaluation (\textbf{Q1.1})}\label{q1.1}
% \textcolor{red}{To answer \textbf{Q1.1} regarding whether incorporating the \textbf{LH}-plugin can lead to significant accuracy improvements in downstream trajectory similarity retrieval task, we conducted experiments in two parts: spatial trajectories and spatio-temporal trajectories.}
\noindent\textbf{Spatial Trajectory.} To validate the enhancement of the \textbf{LH}-plugin on spatial trajectory similarity over the original models, we compare the accuracy of the models before and after integrating the \textbf{LH}-plugin, demonstrating its effectiveness in trajectory similarity learning. The results are presented in Table \ref{tbl:acc}, where the hit rates are displayed as percentages. The results unequivocally demonstrate that:

\noindent(1) Almost universally across all spatial trajectory similarity methods, datasets, and models, there has been a significant improvement in hit rate accuracy after integrating the \textbf{LH}-plugin. Particularly noteworthy are \textbf{Neutraj} and \textbf{Traj2SimVec}, where, in the Chengdu dataset, both methods saw an approximate 30\% increase in accuracy on EDR distance for instance. This clearly demonstrates that the \textbf{LH}-plugin effectively addresses the triangle inequality problem in spatial trajectory similarity representation learning, thereby enhancing the model's effectiveness in similarity computation.

\noindent(2) Models with higher original precision demonstrate a more substantial proportional improvement after integrating the \textbf{LH}-plugin (e.g., the DTW results under Neutraj and the SSPD results under Traj2SimVec). This improvement is attributed to the original models already nearing their performance limits in Euclidean space. Thus, the introduction of hyperbolic space significantly enhances the models' precision. Conversely, minimal improvements are observed in models with relatively low initial precision, such as \textbf{TrajGAT}'s calculation of EDR on the Porto dataset. In such instances, where the original precision is particularly low, the influence of the triangle inequality constraint on similarity accuracy is less pronounced, resulting in less significant precision gains post \textbf{LH}-plugin integration.
% Thus, the introduction of hyperbolic space can significantly enhance the precision of the models. Conversely, we also observed minimal improvement  where its original precision is relateively low, (e.g. \textbf{TrajGAT}'s calculation of EDR distance on the Porto dataset). In these cases when the original precision is particularly low, the constraint of the triangle inequality exerts less influence on similarity accuracy, leading to less significant precision gains after incorporating the \textbf{LH}-plugin.}
This observation is consistent with Figure~\ref{fig:tri_str}, where it is stated that \textit{the triangle inequality contradiction has a more pronounced impact on methods with higher precision}.

\noindent(3) Cross-comparison among various similarity methods indicates that the DTW distance exhibits the most significant improvement after integrating the \textbf{LH}-plugin. This enhancement suggests that DTW distance is particularly impacted by the triangle inequality constraint.

\begin{table}
\vspace{0.05in}
\setlength{\tabcolsep}{1pt}
\centering
\caption{\textcolor{black}{Accuracy results of ST2Vec and Tedj}}
\vspace{-0.09in}
\label{tbl:acc_st2vec}
\begin{adjustbox}{center} 
\resizebox{1\linewidth}{!}{
\begin{tabular}{l|l|cccc|cccc}
\toprule
\toprule
&& \multicolumn{4}{c|}{ST2Vec}&\multicolumn{4}{c}{Tedj}
\\ \hline
sim & plugin & HR@5 & HR@10 & HR@50  & NDCG50 & HR@5 & HR@10 & HR@50  & NDCG50 \\
\hline
\multirow{3}{*}{TP} & Original & 40.21 & 46.61 & 58.54  & 0.5952 & 5.79 & 8.04 & 20.03 & 0.1870 \\
& \textbf{LH}-plugin & 47.46 & 53.62 & 64.32 & 0.6578 & 6.29 & 8.73 & 21.70 & 0.2025 \\
& \%Increase & +18.0\% & +15.0\% & +9.9\% & +10.5\% & +8.6\% & +8.6\% & +8.3\% & +8.3\%\\
\hline
\multirow{3}{*}{DITA} & Original & 40.96 & 46.76 & 54.72 & 0.5939 & 6.44 & 8.83 & 21.22 & 0.1953 \\
& \textbf{LH}-plugin & 49.06 & 54.30 & 62.03 & 0.6877 & 6.91 & 9.53 & 23.98 & 0.2203 \\
& \%Increase & +19.8\% & +16.1\% & +13.4\% & +15.8\% & +7.3\% & +7.9\% & +13.0\% & +12.8\%\\

\hline 
\multirow{3}{*}{\shortstack[l]{discret \\ fréchet}} & Original & 42.50 & 49.52 & 61.97 & 0.6310& 6.23 & 8.62 & 20.23 & 0.1848 \\
& \textbf{LH}-plugin & 46.72 & 53.02 & 64.74 & 0.6712 & 6.67 & 9.23 & 21.98 & 0.2096 \\
& \%Increase & +9.9\% & +7.1\% & +4.5\% & +6.4\% & +7.1\% & +7.1\% & +8.6\% & +13.4\% \\

\bottomrule
\bottomrule
\end{tabular}
}
\end{adjustbox}
\vspace{-0.25in}
\end{table}

% \marginpar{\vspace{3.6cm}\scriptsize{\textcolor{blue}{R2.O2}}}
\noindent\textbf{Spatio-temporal Trajectory.} To complement the evaluation of \textbf{LH}-plugin's performance on spatio-temporal trajectories, we contrast the results of the spatio-temporal similarity computation model \textbf{ST2Vec} and \textbf{Tedj} before and after the addition of \textbf{LH}-plugin, as shown in Table \ref{tbl:acc_st2vec}. It is evident that:

\noindent(1) After integrating the \textbf{LH}-plugin, \textbf{ST2Vec} shows significant improvements in calculations of TP, DITA, and discrete Fréchet distances, with DITA distance notably improving by nearly 20\%. 
\textcolor{black}{(2) After integrating the \textbf{LH}-plugin, \textbf{Tedj} achieved a stable accuracy improvement of around 7\% in various similarity measures.}
% \marginpar{\vspace{-4.6cm}\scriptsize{\textcolor{blue}{R2.O2}}}
These results clearly demonstrates \textbf{LH}-plugin's capability in addressing the triangle inequality constraints inherent in spatio-temporal trajectory similarity learning. It also indirectly highlights that triangle inequality constraints indeed pose a bottleneck for similarity computations in representation learning models.

\vspace{-0.05in}
\subsection{Triangle Inequality Violation Evaluation (\textbf{Q1.2})}
\label{q1.2}
\vspace{-0.03in}

\begin{figure} % [h!] 帮助确定图片的精确位置
\centering % 使图片居中显示
\includegraphics[width=0.93\linewidth]{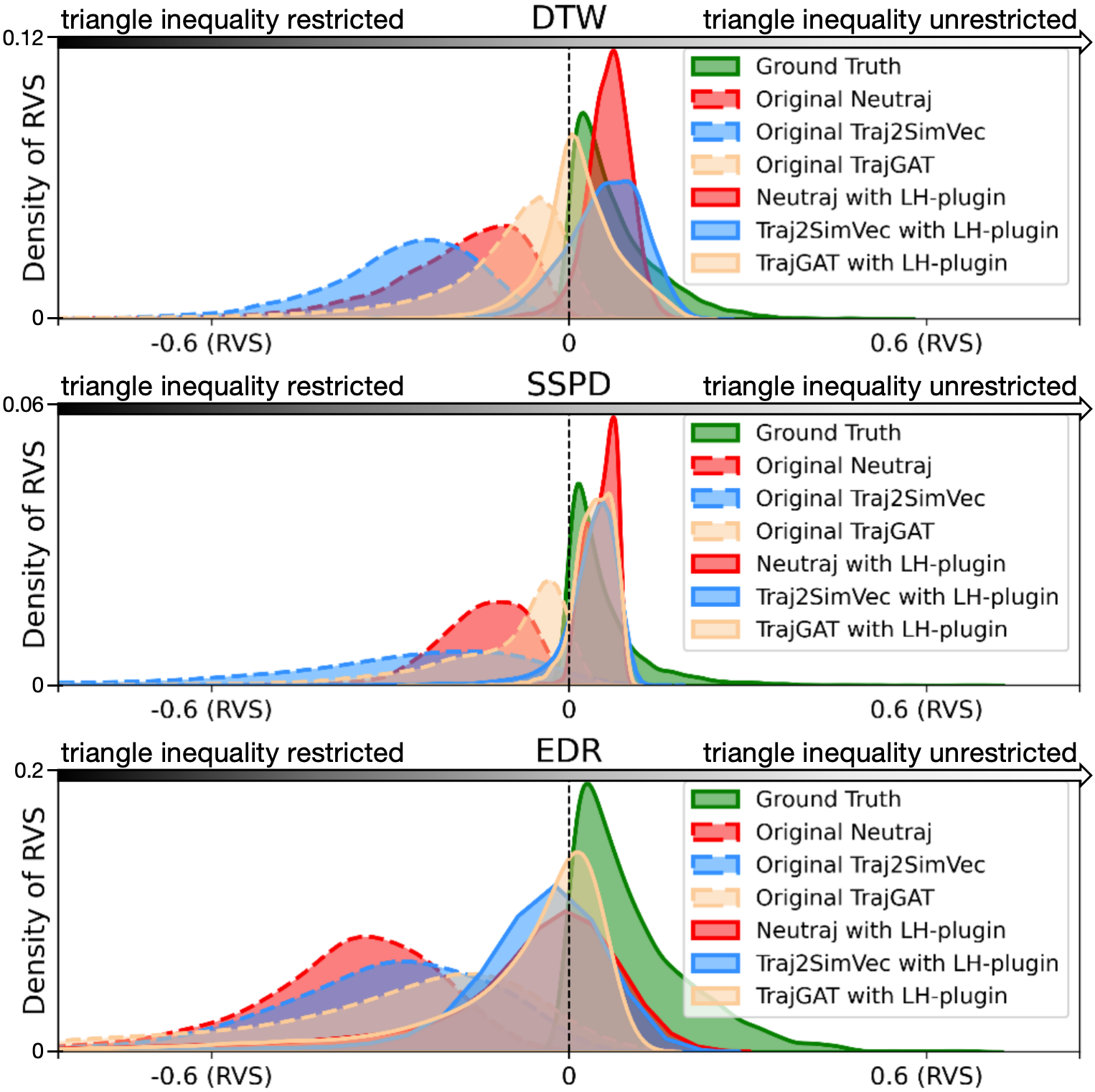} % 设置图片宽度为文本宽度的80%
\vspace{-0.05in}
\caption{\textcolor{black}{RVS Distribution Comparison: Capabilities of Unconstraint from the Triangle Inequality}} % 图片的标题
\vspace{-0.36in}
\label{fig:prof} % 用于文内引用的标签
\end{figure}

% The previous experiments have demonstrated that incorporating the \textbf{LH}-plugin indeed enhances the accuracy of the original trajectory similarity representation learning models. However, it remains unverified whether this improvement is a result of resolving the triangle inequality constraint of the original models. \textcolor{red}{To answer \textbf{Q1.2} in this subsection, we will empirically validate whether the inclusion of \textbf{LH}-plugin alleviates the triangle inequality constraint issue inherent in the original models.}

In this section, we empirically validate whether the inclusion of \textbf{LH}-plugin alleviates the triangle inequality constraint issue inherent in the original models.
Specifically, we randomly select 1 million trajectory triplets from the Chengdu dataset that exhibit conflicts between similarity metrics and the triangle inequality (to facilitate the observation of results, non-conflicting triplets are not included to avoid diluting the outcomes; the proportion of conflicts can be seen in Table \ref{tbl:constraint_dataset} under the \textbf{ARVS}). We record the Euclidean distances in the original model and the Fusion Distances in the model with the \textbf{LH}-plugin. Then, we calculate the \textbf{RVS} (defined in Section \ref{subsec:triag_vio}) for the Similarity Ground Truth, Euclidean distance, and Fusion distance of these triplets respectively.

\noindent \textit{Remark.} For the selected triplets that violate the triangle inequality, the Ground Truth \textbf{RVS} values should all exceed zero, indicating that the metric encounters a conflict with such triplets, where higher values signify a more pronounced conflict. In terms of model learning using Euclidean or Fusion distances, a positive \textbf{RVS} demonstrates that the model can generate distance triples that are not restricted by the triangle inequality. Conversely, a negative \textbf{RVS} indicates adherence to the triangle inequality constraint. Furthermore, the closer the distribution of \textbf{RVS} values aligns with the Ground Truth, the greater the precision achieved by the model.
% \marginpar{\vspace{-5cm}\scriptsize{\textcolor{blue}{R1.O1}}}

% For the selected triplets that violate the triangle inequality, the Ground Truth \textbf{RVS} should all be greater than 0, which means such metric has a conflict on such triplets, with higher values denoting a more pronounced conflict. 
% For model learning Euclidean distance or Fusion distance, a positive \textbf{RVS} indicates the model can generate distance triples unrestricted by the triangle inequality, whereas a negative \textbf{RVS} signifies compliance with the triangle inequality constraint. Additionally, the closer the distribution of \textbf{RVS} values to the Ground Truth, the better the precision achieved. 

Here, we present the probability density of \textbf{RVS} under different methods, as shown in Figure \ref{fig:prof}, from which we draw the following conclusions:

\noindent(1) The Euclidean embeddings generated by the original models (i.e., the part of the dashed line) fall almost entirely on the negative half-axis, representing the original models are strictly constrained by the triangle inequality. However, the Ground Truth values are almost all located on the positive half-axis, such conflicts of triangle inequality constraint in DTW, SSPD, and EDR significantly impact the learning accuracy in the Euclidean space. 
% \marginpar{\vspace{-5cm}\scriptsize{\textcolor{blue}{R1.O1}}}

\noindent(2) After incorporating the \textbf{LH}-plugin, the distribution of \textbf{RVS} in different models shows a trend towards the positive half-axis, thus becoming closer to the ground truth. This suggests that the inclusion of the \textbf{LH}-plugin indeed addresses the issue of the triangle inequality constraint in representation learning. For instance, considering the \textbf{Neutraj} method in the context of DTW distance, the negative instances in its \textbf{RVS} distribution nearly disappear after the integration of the \textbf{LH}-plugin, which corresponds to the substantial improvement observed in its accuracy results (HR@5 increased from 53.54\% to 68.33\%, as seen in Table \ref{tbl:acc}).

\begin{figure} % [h!] 帮助确定图片的精确位置
\vspace{-0.06in}
\centering % 使图片居中显示
\includegraphics[width=0.8\linewidth]{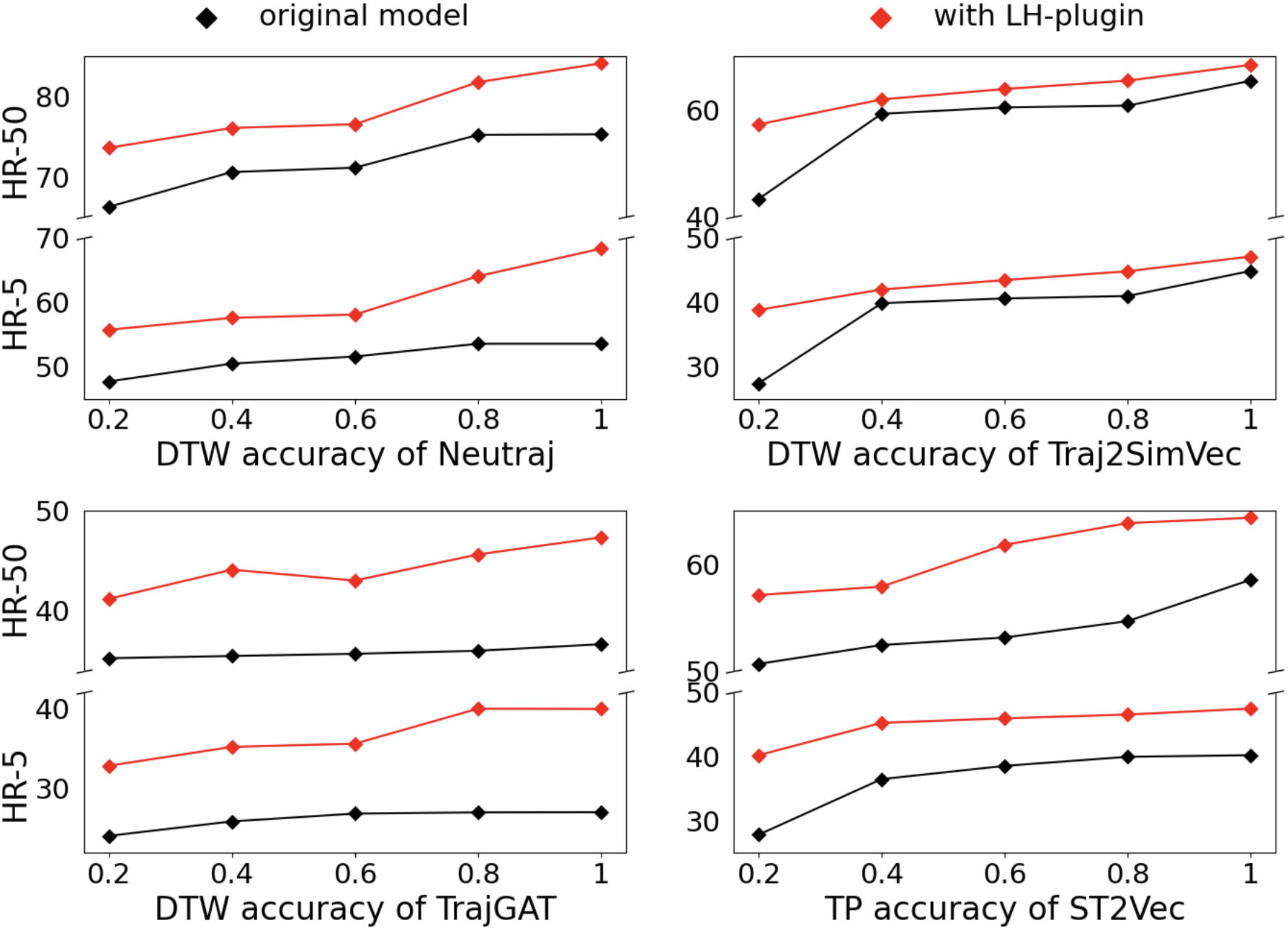} 
\vspace{-0.05in}
\caption{\textcolor{black}{Scalability Comparison of \textbf{LH}-plugin}} 
\label{fig:scala}
\vspace{-0.27in}
\end{figure}

\vspace{-0.07in}
\subsection{Efficiency Evaluation (\textbf{Q2.1})}
\label{q2.1}
\vspace{-0.05in}
% \begin{table*}[h]
% \centering
% \caption{Comparison of Model Performance with and without LH-plugin}
% \label{tab:model_performance}
% \scalebox{1}{
% \begin{tabular}{@{}l|cc|cc|cc|cc@{}}
% \toprule
% \toprule
% & \multicolumn{2}{c|}{TrajGAT} & \multicolumn{2}{c|}{Neutraj} & \multicolumn{2}{c|}{Traj2SimVec} & \multicolumn{2}{c}{ST2Vec} \\ 
% \cline{2-9}
% & Original & with LH-plugin & Original & with LH-plugin & Original & with LH-plugin & Original & with LH-plugin \\ \hline
% Training Time (min)    & 22.31    & 24.35  & 14.17        & 16.27         & 1.32     & 2.62           & 3.01     & 6.28                        \\
% Training GPU (GB)     & 7.39   & 8.66      & 2.04        & 2.72       & 3.25   & 3.73         & 3.93   & 4.33                  \\
% Inference Time (ms) & 11.96        & 12.30              & 1.83       & 2.06               & 0.92        & 1.15              & 1.71        & 1.97              \\

% Model Size (MB)          & 36.40      & 40.12            & 131.40    & 132.87          & 68.03     & 71.76           & 14.09     & 15.16           \\\bottomrule \bottomrule
% \end{tabular}
% }

% \end{table*}

\begin{table*}
\vspace{-0.02in}
\setlength{\tabcolsep}{1.3pt}
\centering
\caption{\textcolor{black}{The Additional Consumption Introduced by \textbf{LH}-plugin in Trajectory Retrieval}}
\vspace{-0.08in}
\label{tbl:effi}
\scalebox{0.65}{
\begin{tabular}{@{}c|ccc|ccc|ccc|ccc@{}}
\toprule
\toprule
\multirow{2}{*}{\shortstack{Trajectory\\ number}} & \multicolumn{3}{c|}{TrajGAT--(time/memory)} & \multicolumn{3}{c|}{Neutraj--(time/memory)} & \multicolumn{3}{c|}{Traj2SimVec--(time/memory)} & \multicolumn{3}{c}{ST2Vec--(time/memory)} \\ 
\cline{2-13}
& Original & with LH-plugin & \%Increase  & Original & with LH-plugin & \%Increase & Original & with LH-plugin & \%Increase & Original & with LH-plugin & \%Increase \\ \hline
10k   & 0.45s/84MB        & 0.46s/94MB        & 1.8\%/12.5\% & 0.21s/274MB        & 0.21s/279MB       & 2.63\%/2.15\% & 0.19s/147MB        & 0.19s/156MB        & 2.68\%/7.07\% & 0.21s/39MB        & 0.21s/44MB        & 2.97\%/12.76\% \\
100k  & 1.86s/186MB        & 1.87/201MB        & 0.43\%/8.31\% & 1.63s/375MB       & 1.63s/387MB        & 0.34\%/2.9\% & 1.61s/249MB        & 1.61s/264MB       & 0.31\%/6.18\% & 1.63s/141MB        & 1.63s/151MB       & 0.38\%/7.08\% \\
1m & 15.60s/1.07GB       & 15.61s/1.15GB      & 0.05\%/7.85\% & 15.34s/1.27GB    & 15.35s/1.35GB       & 0.04\%/6.31\% & 15.34s/1.14GB       & 15.34s/1.22GB       & 0.03\%/7.41\% & 15.33s/1.03GB       & 15.34s/1.11GB       & 0.04\%/7.66\% \\ \bottomrule
\bottomrule
\end{tabular}}
\vspace{-0.2in}
\end{table*}

In this section, we assess whether the \textbf{LH}-plugin significantly increases inference latency \textcolor{black}{ and memory consumption}, potentially impacting the trajectory retrieval process. The integration of new modules into existing representation learning frameworks is likely to extend the inference time. Our goal is to determine if the trajectory retrieval efficiency becomes unacceptably compromised after incorporating the \textbf{LH}-plugin compared to the baseline models. Prior studies indicate that trajectory embedding methods generally outperform non-learning-based rule optimization approaches (e.g., \cite{dtw_para, dtw_prun}) due to the ability to pre-embed a large number of trajectories offline. We conduct our efficiency evaluation based on this pre-embedding strategy for trajectories, comparing the end-to-end latency of online similar trajectory retrieval before and after the integration of the \textbf{LH}-plugin. As detailed in Table~\ref{tbl:effi}, the number of trajectories indicates the size of the trajectory database. While integrating the \textbf{LH}-plugin does result in increased retrieval latency, this increment is marginal—less than 0.05\%—especially in real-world datasets that scale to millions of trajectories, rendering it negligible in comparison to the overall retrieval time. \textcolor{black}{(From the sizes of commonly used trajectory databases shown in Table I, trajectory datasets for commercial use typically exceed millions of entries.)} \textcolor{black}{Similarly, the additional memory overhead brought by \textbf{LH}-plugin is clearly not worth mentioning. The increase in memory overhead is generally less than 8\%.}

\vspace{-0.06in}
\subsection{Scalability Evaluation (\textbf{Q2.2})}
\label{q2.2}
\vspace{-0.04in}

% \textcolor{red}{In this section, we answer the question \textbf{Q2.2}, examining whether the introduction of the \textbf{LH}-plugin affects the scalability of the original models.}
% In this section, we aim to verify the differential effects of the \textbf{LH}-plugin on different datasets' scalability. 
In this section, we aim to examine whether the introduction of the \textbf{LH}-plugin affects the scalability of the original models.
We compare the impact of different training data sizes on accuracy. Specifically, we sample 20\%, 40\%, 60\%, 80\%, and 100\% from the Chengdu dataset (TDrive for \textbf{ST2Vec}) to train different trajectory similarity models. The results, as depicted in Figure \ref{fig:scala}, lead us to conclude: 

\noindent(1) With an increase in training data, there is a significant improvement in trajectory similarity accuracy across different models. 

\noindent(2) The inclusion of the \textbf{LH}-plugin results in noticeable enhancements in accuracy across all data sizes, with the accuracy curves before and after adding the \textbf{LH}-plugin exhibiting similar shapes. Thus, it can be concluded that the \textbf{LH}-plugin can significantly improve the accuracy of trajectory similarity models regardless of the training data size.

\vspace{-0.07in}
\subsection{Robustness Evaluation (\textbf{Q3})}
\label{q2.3}
\vspace{-0.04in}

\begin{figure} % [h!] 帮助确定图片的精确位置
\vspace{-0.04in}
\centering
\includegraphics[width=0.75\linewidth]{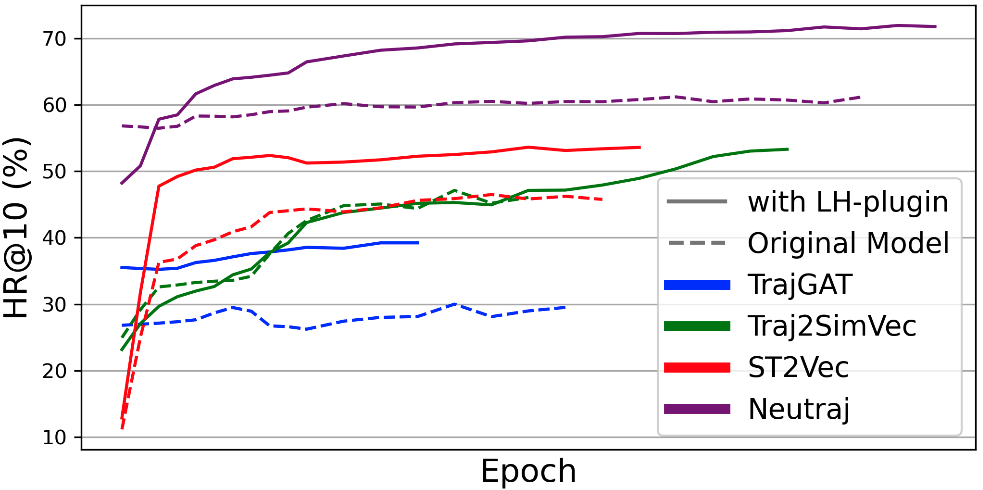}
\vspace{-0.08in}
\caption{Robustness Comparison of \textbf{LH}-plugin} 
\label{fig:stab} 
\vspace{-0.25in}
\end{figure}

% \textcolor{red}{In this section, we primarily answer Question \textbf{Q2.3}: Does the introduction of the \textbf{LH}-plugin (hyperbolic space) affect the convergence of the original model (Euclidean space) during training?} 
In this section, we aim to evaluate whether the \textbf{LH}-plugin is robust to enhance the similarity retrieval process.
In the aforementioned experiments, we observe an intriguing phenomenon: the training curves with the \textbf{LH}-plugin are more stable compared to the original model. If there are significant fluctuations in accuracy in the original model, the inclusion of the \textbf{LH}-plugin results in a smoother curve. We illustrate this with the training curves using DTW distance as an example in Figure \ref{fig:stab}, thus we get the following conclusion:

\noindent (1) For the same original model, the smoother accuracy curve with the \textbf{LH}-plugin helps avoid premature convergence due to fluctuations, allowing for more thorough training. 

\noindent (2) Additionally, the smoother accuracy curve with the \textbf{LH}-plugin brings another advantage: rapid convergence after reaching a plateau, which is particularly pronounced in the \textbf{TrajGAT} model.

\vspace{-0.06in}
\subsection{Ablation Evaluation (\textbf{Q4})}
\label{q3}
\vspace{-0.04in}

% \textcolor{red}{In this section, we mainly address question \textbf{Q3}, can the effectiveness of our principal contributions in \textbf{LH}-plugin be confirmed quantitatively.} 
In this section, we aim to quantitatively to validate the effectiveness of our three principal contributions: \textbf{1) the Lorentz distance in Hyperbolic space}, \textbf{2) the Cosh Projection to Hyperbolic space}, and \textbf{3) the Dynamic Fusion Distance}, we structure the \textbf{LH}-plugin around these elements and conduct a series of incremental ablation experiments. The detailed results are presented in Table \ref{tab:abla}. Below, we discuss the ablation design and findings corresponding to these three key aspects:
% \marginpar{\vspace{-4cm}\scriptsize{\textcolor{blue}{R1.O2}}}

\noindent\textbf{(1) The Lorentz Distance in Hyperbolic Space:}
We assess whether integrating the Lorentz distance within hyperbolic space could substantially improve model accuracy by overcoming the limitations posed by the triangle inequality. This involves comparing the accuracy of the traditional Euclidean space representation learning (``original'' row in Table \ref{tab:abla}) with that of the hyperbolic space representation using the Lorentz distance via Vanilla Projection (``lh-vanilla'' row in Table \ref{tab:abla}). The findings indicate a notable increase in similarity accuracy across most metrics and models when switching to the Lorentz distance. Therefore, the substantial improvements noted in most models underscore the efficacy of the Lorentz metric in capturing deeper relational nuances that Euclidean metrics often overlook. However, the marginal improvements in some models suggest that the integration of hyperbolic space may require careful calibration to fully realize its potential, particularly in preserving semantic integrity during projection.

\noindent\textbf{(2) The Cosh Projection to Hyperbolic Space:}
This part of the study aims to determine whether employing the Cosh Projection instead of the Vanilla Projection for mapping Euclidean embeddings into hyperbolic space could address the issue of Lorentz distance collapsing to zero with large-norm Euclidean embeddings (as outlined in Section 4.1), thereby significantly enhancing model accuracy. We compare the accuracy differences between the Vanilla Projection (``lh-vanilla'' in Table \ref{tab:abla}) and the Cosh Projection (``lh-cosh'' in Table \ref{tab:abla}). The results decisively demonstrate that switching to Cosh Projection substantially improved similarity accuracy. Therefore, the Cosh Projection emerges as a crucial enhancement over the Vanilla Projection method, particularly in addressing the challenge of distance diminishment with increasing Euclidean norms. This projection method stabilizes the expansion of trajectory embeddings into hyperbolic space, thereby maintaining consistent representational fidelity across varying magnitudes of embeddings. The pronounced improvements with the ``lh-cosh'' setup in our experiments demonstrate the critical role of this method in maintaining robust distance metrics, even under scaling challenges posed by larger-norm embeddings.

\noindent\textbf{(3) The Dynamic Fusion Distance:}
Finally, to explore whether integrating the Dynamic Fusion Distance enhances model accuracy, we combine the Lorentz distance obtained via Cosh Projection with the conventional Euclidean distance using the Dynamic Fusion Module (described in Section \ref{sec:Fusion_dist}). This configuration represents the comprehensive application of the \textbf{LH}-plugin (``fusion-dist'' in Table \ref{tab:abla}). When compared with results from solely using Euclidean distance (``original'' in Table \ref{tab:abla}) and only Lorentz distance (``lh-cosh'' in Table \ref{tab:abla}), the integration of Dynamic Fusion Distance demonstrates significant accuracy improvements across various models and metrics, validating the efficacy of this approach. The fusion approach not only accommodates the variability in triangle inequality constraints but also enhances the model's flexibility to accurately reflect the true similarity across diverse trajectory datasets. 
% \marginpar{\vspace{-4.2cm}\scriptsize{\textcolor{blue}{R1.O1}}}

\begin{figure}
\vspace{0.03in}
\centering
\includegraphics[width=1\linewidth]{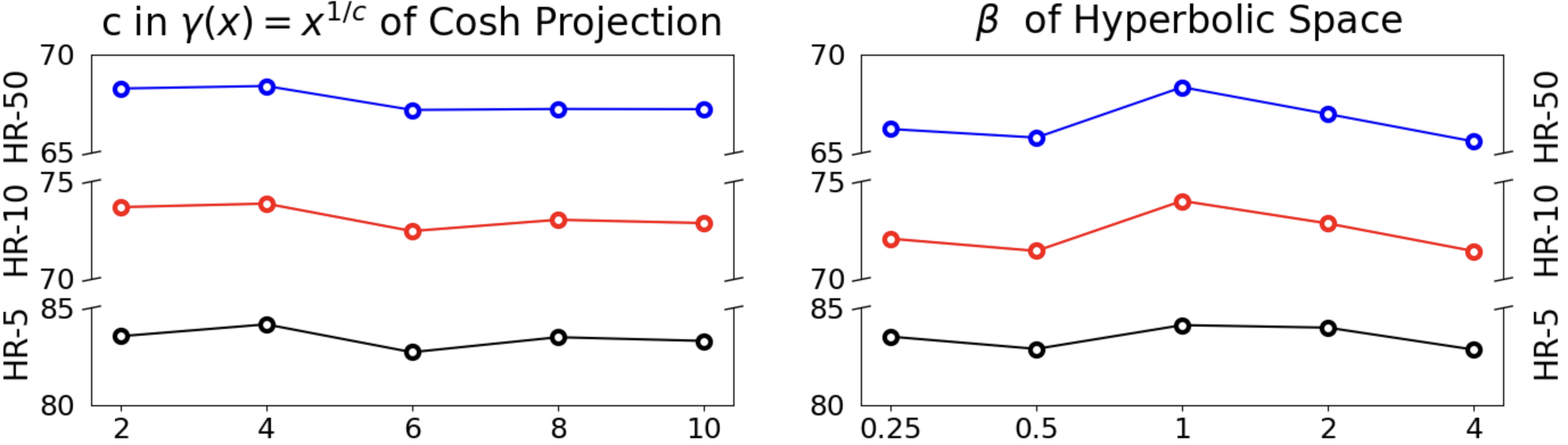}
\vspace{-0.22in}
\caption{\textcolor{black}{Hyper-parameter Evaluation}} % 图片的标题
\label{fig:hyper}
\vspace{-0.27in}
\end{figure}

\begin{table*}[t]
\vspace{-0.03in}
\centering
\caption{Ablation Evaluation}
\vspace{-0.08in}
\label{tab:abla}
\setlength{\tabcolsep}{4.2pt}
\scalebox{0.75}{
\begin{tabular}{c|c|cccc|cccc|cccc}
\toprule
\toprule
\multicolumn{2}{c|}{} & \multicolumn{4}{c|}{DTW} & \multicolumn{4}{c|}{SSPD} & \multicolumn{4}{c}{EDR} \\
\cline{3-14}
\multicolumn{2}{c|}{} & original & lh-vanilla & lh-cosh & fusion-dist & original & lh-vanilla & lh-cosh & fusion-dist & original & lh-vanilla & lh-cosh & fusion-dist \\
\hline
\multirow{3}{*}{Neutraj}& HR@5 & 53.54 & 65.03 & 67.07 & 68.33 & 49.80 & 50.05 & 57.66 & 58.25 & 8.81 & 8.07 & 11.87 & 12.06 \\
&HR@10 & 60.36 & 71.67 & 73.54 & 73.99 & 54.48 & 55.18 & 63.03 & 65.72 & 10.71 & 10.79 & 13.51 & 14.46 \\
&HR@50  & 75.28 & 83.64 & 83.87 & 84.11 & 66.81 & 67.52 & 78.56 & 80.32 & 13.21 & 14.67 & 15.89 & 17.48 \\
\hline
\multirow{3}{*}{TrajGAT} & HR@5  & 27.07 & 36.85 & 39.11 & 39.92 & 38.61 & 40.78 & 42.55 & 45.10 & 5.74 & 6.02 & 6.51 & 6.69 \\
&HR@10 & 30.31 & 38.96 & 41.79 & 41.95 & 43.73 & 46.44 & 48.89 & 50.33 & 6.63 & 6.72 & 7.05 & 7.49 \\
&HR@50 & 36.69 & 42.31 & 47.06 & 47.32 & 55.37 & 58.11 & 60.81 & 61.74 & 8.19 & 8.15 & 8.33 & 8.65 \\
\hline
\multirow{3}{*}{Traj2SimVec} & HR@5  & 44.87 & 45.66 & 46.49 & 47.11 & 47.68 & 51.61 & 56.02 & 57.37 & 11.90 & 11.98 & 13.45 & 15.89 \\
& HR@10 & 50.38 & 51.23 & 51.99 & 53.30 & 52.64 & 55.73 & 60.35 & 62.99 & 14.51 & 15.11 & 16.72 & 19.68 \\
& HR@50 & 65.41 & 65.97 & 66.89 & 68.40 & 66.05 & 68.12 & 71.86 & 74.04 & 17.56 & 17.99 & 20.14 & 25.08 \\
\bottomrule
\bottomrule
\end{tabular}
}
\vspace{-0.24in}
\end{table*}

\vspace{-0.07in}
\subsection{Hyper-Parameter Evaluation (\textbf{Q5})}\label{q5}
\vspace{-0.04in}

In the \textbf{LH}-plugin, we have set the following two adjustable hyper-parameters, $\beta$ and $c$. 

The parameter $\beta$ is a hyper-parameter in the hyperbolic space $\mathcal{H}(\beta)=\{\mathbf{a}\in\mathbb{R}^{2}:\ \langle\mathbf{a},\mathbf{a}\rangle = -\beta,\ a_0 \geq \sqrt{\beta}\}$, which controls the shape of the hyperbolic space. This parameter significantly influences the shape of the hyperbolic space near the origin, thereby affecting the accuracy of Lorentz distance when the norm of Euclidean embedding is small. 

The parameter $c$ in $\gamma(x) = x^{1/c}$, as mentioned in Section \ref{gamma}, is a non-linear function introduced to adjust the norm of Euclidean embeddings. When $c$ is set to a larger value, it can alleviate the issue of large norms in Euclidean space to some extent, but an excessively large $c$ might lead to a loss of digital accuracy due to its actual implementation using finite Taylor series expansion. 

The experimental results for these two hyper-parameters are shown in the Figure \ref{fig:hyper}. Based on these results, we ultimately selected $c=4$ and $\beta=1$.

\vspace{-0.05in}
\section{Related Work}
\vspace{-0.06in}
\subsection{Trajectory Representation Learning for Similarity Computation}
\vspace{-0.04in}

Trajectory similarity is a fundamental area of research within trajectory data management and serves as a critical precursor to various downstream tasks. 
The computational complexity of traditional distance functions~\cite{dtw, erp, edr, hausdorff, frechet, sspd} —is super-quadratic during both the computation and retrieval phases. 
Particularly in the scenarios involving similarity searches, it is necessary to perform pairwise similarity calculations for every trajectory in the database, which entails exhaustive pairwise comparisons between each pair of points in the trajectories.

The\, introduction\, of\, trajectory\, representation\, learning\, for 

\noindent similarity computation has advanced the field by vectorizing trajectories, allowing the use of vector distances to measure similarity between trajectories. This approach addresses the combinatorial complexity encountered during retrieval and computation. A pioneering work in this field is Li et al.~\cite{pioneer}, which first proposed a framework that uses deep learning networks to vectorize variable-length trajectories. It measures the similarity between trajectories by distances among vectors, employs methods like locality-sensitive hashing for preliminary screening of approximate trajectories, and subsequently computes the actual similarity distances. 
Further advancements were made by Yao et al.~\cite{neutraj}, which improved upon earlier efforts by splitting the trajectory space into grids and aggregating neighboring information. 
% In addition, this work used a GRU network that incorporates node proximity relationships into the trajectory representation vectors.
However, they did not capitalize on the information of sub-trajectories on the trajectory. To address this issue, Zhang et al.~\cite{traj2s} introduced sub-trajectory information to enhance semantic similarity between trajectories based on sub-trajectory similarities. Nevertheless, the trajectory lengths processed by these models are often much shorter than those encountered in real-world data.
Therefore, Yao et al.~\cite{TrajGAT} focused on 
% addressing 
long trajectory issues in real-world data by employing quadtree structured encoding of trajectory space and utilizing graph attention networks for trajectory embedding. 
Another approach to this issue involves simplifying trajectories by incorporating temporal dimension information. For example, Fang et al.~\cite{st2vec} incorporated the time dimension into similarity representation learning, integrating time and space dimensions through co-attention mechanisms to jointly compute similarity. \textcolor{black}{Tedjopurnomo et al.~\cite{tedj} introduced the concept of spatio-temporal grids to capture the spatio-temporal relationships in spatio-temporal trajectories.}
% \marginpar{\vspace{-1cm}\scriptsize{\textcolor{blue}{R2.O2}}}
\vspace{-0.01in}

Despite these advances, nearly all research in this field has focused on refining network models to map trajectories into Euclidean space. However, \textit{a significant gap remains in addressing the inherent contradictions between Euclidean space and similarity metrics, such as the triangle inequality, which have yet to be thoroughly explored.}

\vspace{-0.05in}
\subsection{Representation Learning in Hyperbolic Space}
\vspace{-0.04in}

Hyperbolic space possesses a natural advantage in embedding representations with high semantic richness~\cite{Hypergroup}, particularly excelling at handling hierarchical dependencies such as tree-like structures. By hyperbolic space, extensive and intricate dependencies in the real world can be decomposed into tree-shaped forms, due to their potential tree-like structural features~\cite{treelike, hierar}. Inspired by this finding, the study of learning hyperbolic embeddings has garnered significant attention in various domains of representation learning, primarily 
 dividing into two orientations:

In the first field, some research endeavors have attempted to reconstruct existing neural network models within hyperbolic space to achieve end-to-end fully hyperbolic neural network representation learning. Ganea et al.~\cite{hnn}, pioneers in this area, employ the formalism of Möbius gyrovector spaces with Riemannian geometry to derive various deep learning architectures in hyperbolic space, such as multinomial logistic regression and feed-forward networks. 
% Chami et al.~\cite{hgcn} extended this work by proposing the hyperbolic graph neural network, the hyperbolic space form of graph neural networks. 
However, due to the fact that GPU operators are not optimized for computations related to hyperbolic space, the training and inference process of hyperbolic neural networks are evidently inefficient. Additionally, migrating existing representation learning methods from Euclidean space to hyperbolic space may have unintended side effects on their effectiveness.

Therefore, the other orientation of research opts for a hybrid network structure instead of a purely hyperbolic neural network pipeline, only in the final layer or a specific part, introducing the hyperbolic model to map Euclidean space embeddings into hyperbolic space. These approaches~\cite{hyper-vision, hyper-learning, hyper-img}, much similar to ours, using the hybrid network and resulting in an incremental plugin, can be utilized to improve existing Euclidean space methods. 
% As a plugin instance, Ermolov et al.~\cite{hyper-vision} proposed a loss based on the cross-entropy of hyperbolic distance for Vision Transformers to capture hierarchical relationships among images. Yan et al.~\cite{hyper-learning} incorporated hyperbolic-based clustering in the final layer of the representation learning pipeline, achieving exceptional results in unsupervised tasks.
In this orientation, \textit{previous research has rarely directly addressed the intuitive conflicts between representation learning and metrics space, such as triangle inequality.}

\vspace{-0.02in}
\section{Conclusion}
\vspace{-0.03in}
In this paper, we conduct an in-depth study of the triangle inequality issue in trajectory similarity representation learning, specifically the conflict between the triangle inequality constraint in Euclidean space and trajectory similarity metrics. 
To address this, we propose the \textbf{LH}-plugin, a model-agnostic plugin for any trajectory similarity representation learning model. This plugin utilizes the Lorentz distance in hyperbolic space to circumvent the triangle inequality constraint present in Euclidean space. It employs the Cosh function to optimize the projection from Euclidean space to hyperbolic space and dynamically adjusts the fusion of Lorentz distance and Euclidean distance based on different trajectory combinations using fusion distance. Extensive experiments were conducted on real-world datasets using various mainstream trajectory similarity representation learning models, validating the effectiveness of the \textbf{LH}-plugin.

\clearpage
\bibliographystyle{IEEEtran}
\bibliography{sample}

% Generated by IEEEtran.bst, version: 1.14 (2015/08/26)
\begin{thebibliography}{10}
\providecommand{\url}[1]{#1}
\csname url@samestyle\endcsname
\providecommand{\newblock}{\relax}
\providecommand{\bibinfo}[2]{#2}
\providecommand{\BIBentrySTDinterwordspacing}{\spaceskip=0pt\relax}
\providecommand{\BIBentryALTinterwordstretchfactor}{4}
\providecommand{\BIBentryALTinterwordspacing}{\spaceskip=\fontdimen2\font plus
\BIBentryALTinterwordstretchfactor\fontdimen3\font minus \fontdimen4\font\relax}
\providecommand{\BIBforeignlanguage}[2]{{%
\expandafter\ifx\csname l@#1\endcsname\relax
\typeout{** WARNING: IEEEtran.bst: No hyphenation pattern has been}%
\typeout{** loaded for the language `#1'. Using the pattern for}%
\typeout{** the default language instead.}%
\else
\language=\csname l@#1\endcsname
\fi
#2}}
\providecommand{\BIBdecl}{\relax}
\BIBdecl

\bibitem{survey1}
W.~Chen, Y.~Liang, Y.~Zhu, Y.~Chang, K.~Luo, H.~Wen, L.~Li, Y.~Yu, Q.~Wen, C.~Chen, K.~Zheng, Y.~Gao, X.~Zhou, and Y.~Zheng, ``Deep learning for trajectory data management and mining: {A} survey and beyond,'' \emph{arXiv}, vol. abs/2403.14151, 2024.

\bibitem{survey2}
S.~Wang, Z.~Bao, J.~S. Culpepper, and G.~Cong, ``A survey on trajectory data management, analytics, and learning,'' \emph{CSUR}, vol.~54, pp. 1--36, 2021.

\bibitem{survey3}
C.~L. da~Silva, L.~M. Petry, and V.~Bogorny, ``A survey and comparison of trajectory classification methods,'' in \emph{BRACIS}, 2019, pp. 788--793.

\bibitem{survey4}
Z.~Feng and Y.~Zhu, ``A survey on trajectory data mining: Techniques and applications,'' \emph{IEEE Access}, vol.~4, pp. 2056--2067, 2016.

\bibitem{survey5}
S.~Wang, J.~Cao, and S.~Y. Philip, ``Deep learning for spatio-temporal data mining: A survey,'' \emph{TKDE}, vol.~34, no.~8, pp. 3681--3700, 2020.

\bibitem{yuan2}
H.~Yuan and G.~Li, ``Distributed in-memory trajectory similarity search and join on road network,'' in \emph{ICDE}, 2019, pp. 1262--1273.

\bibitem{yuan11}
Y.~Wang, G.~Li, K.~Li, and H.~Yuan, ``A deep generative model for trajectory modeling and utilization,'' \emph{VLDB}, pp. 973--985, 2022.

\bibitem{traj_cluster}
D.~Yao, C.~Zhang, Z.~Zhu, J.~Huang, and J.~Bi, ``Trajectory clustering via deep representation learning,'' in \emph{IJCNN}, 2017, pp. 3880--3887.

\bibitem{traj_cluster2}
X.~Olive, L.~Basora, B.~Viry, and R.~Alligier, ``Deep trajectory clustering with autoencoders,'' in \emph{ICRAT}, 2020.

\bibitem{traj_cluster3}
Z.~Fang, Y.~Du, L.~Chen, Y.~Hu, Y.~Gao, and G.~Chen, ``E 2 dtc: An end to end deep trajectory clustering framework via self-training,'' in \emph{ICDE}, 2021, pp. 696--707.

\bibitem{cluster_survey}
J.~Bian, D.~Tian, Y.~Tang, and D.~Tao, ``A survey on trajectory clustering analysis,'' \emph{arXiv}, 2018.

\bibitem{sim_query}
Z.~Fang, S.~Gong, L.~Chen, J.~Xu, Y.~Gao, and C.~S. Jensen, ``Ghost: A general framework for high-performance online similarity queries over distributed trajectory streams,'' \emph{POMD}, pp. 1--25, 2023.

\bibitem{retrieval2}
Y.~Su, D.~Yao, X.~Zhou, Y.~Zhang, Y.~Fan, L.~Bai, and J.~Bi, ``Tripsafe: Retrieving safety-related abnormal trips in real-time with trajectory data,'' in \emph{SIGIR}, 2023, pp. 2446--2450.

\bibitem{yuan4}
J.~Li, H.~Yuan, G.~Cong, H.~M. Kiah, and S.~Zhang, ``{MAST:} towards efficient analytical query processing on point cloud data,'' \emph{SIGMOD}, pp. 52:1--52:27, 2025.

\bibitem{yuan7}
H.~Yuan, ``An effective deep learning model for route travel time estimation on {A} road network,'' in \emph{ACM TURC}, 2023, pp. 125--126.

\bibitem{yuan10}
H.~Yuan, G.~Li, Z.~Bao, and L.~Feng, ``Effective travel time estimation: When historical trajectories over road networks matter,'' in \emph{SIGMOD}, 2020, pp. 2135--2149.

\bibitem{traj_compress}
R.~Song, W.~Sun, B.~Zheng, and Y.~Zheng, ``{PRESS:} {A} novel framework of trajectory compression in road networks,'' \emph{VLDB}, pp. 661--672, 2014.

\bibitem{traffic}
M.~Shaygan, C.~Meese, W.~Li, X.~G. Zhao, and M.~Nejad, ``Traffic prediction using artificial intelligence: Review of recent advances and emerging opportunities,'' \emph{TR-C}, vol. 145, p. 103921, 2022.

\bibitem{cmx}
M.~Chen, H.~Yuan, N.~Jiang, Z.~Bao, and S.~Wang, ``Urban traffic accident risk prediction revisited: Regionality, proximity, similarity and sparsity,'' in \emph{CIKM}, 2024, pp. 281--290.

\bibitem{yuan3}
H.~Yuan, G.~Cong, and G.~Li, ``Nuhuo: An effective estimation model for traffic speed histogram imputation on {A} road network,'' \emph{VLDB}, no.~7, pp. 1605--1617, 2024.

\bibitem{yuan5}
Z.~Zhao, H.~Yuan, N.~Jiang, M.~Chen, N.~Liu, and Z.~Li, ``{STMGF:} an effective spatial-temporal multi-granularity framework for traffic forecasting,'' in \emph{DASFAA}, 2024, pp. 235--245.

\bibitem{yuan6}
H.~Yuan, S.~Wang, Z.~Bao, and S.~Wang, ``Automatic road extraction with multi-source data revisited: Completeness, smoothness and discrimination,'' \emph{VLDB}, pp. 3004--3017, 2023.

\bibitem{traffic2}
H.~Yuan and G.~Li, ``A survey of traffic prediction: from spatio-temporal data to intelligent transportation,'' \emph{DSE}, vol.~6, no.~1, pp. 63--85, 2021.

\bibitem{yuan8}
H.~Yuan, G.~Li, Z.~Bao, and L.~Feng, ``An effective joint prediction model for travel demands and traffic flows,'' in \emph{ICDE}, 2021, pp. 348--359.

\bibitem{yuan9}
H.~Yuan and G.~Li, ``A survey of traffic prediction: from spatio-temporal data to intelligent transportation,'' \emph{Data Sci. Eng.}, pp. 63--85, 2021.

\bibitem{dtw}
B.~Yi, H.~V. Jagadish, and C.~Faloutsos, ``Efficient retrieval of similar time sequences under time warping,'' in \emph{ICDE}, 1998, pp. 201--208.

\bibitem{dtw_prun}
P.~K. Agarwal, K.~Fox, J.~Pan, and R.~Ying, ``Approximating dynamic time warping and edit distance for a pair of point sequences,'' in \emph{SoCG}, vol.~51, 2016, pp. 6:1--6:16.

\bibitem{sspd}
P.~C. Besse, B.~Guillouet, J.~Loubes, and F.~Royer, ``Review and perspective for distance based trajectory clustering,'' \emph{CoRR}, vol. abs/1508.04904, 2015.

\bibitem{edr}
L.~Chen, M.~T. {\"{O}}zsu, and V.~Oria, ``Robust and fast similarity search for moving object trajectories,'' in \emph{SIGMOD}, 2005, pp. 491--502.

\bibitem{hausdorff}
S.~Atev, G.~Miller, and N.~P. Papanikolopoulos, ``Clustering of vehicle trajectories,'' \emph{TITS}, vol.~11, no.~3, pp. 647--657, 2010.

\bibitem{TMN}
P.~Yang, H.~Wang, D.~Lian, Y.~Zhang, L.~Qin, and W.~Zhang, ``{TMN:} trajectory matching networks for predicting similarity,'' in \emph{ICDE}, 2022, pp. 1700--1713.

\bibitem{trajCL}
Y.~Chang, J.~Qi, Y.~Liang, and E.~Tanin, ``Contrastive trajectory similarity learning with dual-feature attention,'' in \emph{ICDE}, 2023, pp. 2933--2945.

\bibitem{TrajGAT}
D.~Yao, H.~Hu, L.~Du, G.~Cong, S.~Han, and J.~Bi, ``Trajgat: {A} graph-based long-term dependency modeling approach for trajectory similarity computation,'' in \emph{KDD}, 2022, pp. 2275--2285.

\bibitem{road1}
S.~B. Yang, J.~Hu, C.~Guo, B.~Yang, and C.~S. Jensen, ``Lightpath: Lightweight and scalable path representation learning,'' in \emph{SIGKDD}, 2023, pp. 2999--3010.

\bibitem{road2}
S.~Zhou, J.~Li, H.~Wang, S.~Shang, and P.~Han, ``Grlstm: trajectory similarity computation with graph-based residual lstm,'' in \emph{AAAI}, vol.~37, no.~4, 2023, pp. 4972--4980.

\bibitem{road3}
P.~Han, J.~Wang, D.~Yao, S.~Shang, and X.~Zhang, ``A graph-based approach for trajectory similarity computation in spatial networks,'' in \emph{KDD}, 2021, pp. 556--564.

\bibitem{eu}
Euclid, \emph{The Elements}.\hskip 1em plus 0.5em minus 0.4em\relax Dover Publications, 1956, originally published ca. 300 B.C.

\bibitem{eu_metric_book}
W.~Rudin, \emph{Principles of Mathematical Analysis}, 3rd~ed.\hskip 1em plus 0.5em minus 0.4em\relax McGraw-Hill, 1976.

\bibitem{min_metric_book}
R.~M. Wald, \emph{General Relativity}, 1st~ed.\hskip 1em plus 0.5em minus 0.4em\relax University of Chicago Press, 1984.

\bibitem{Colla}
C.-K. Hsieh, L.~Yang, Y.~Cui, T.-Y. Lin, S.~Belongie, and D.~Estrin, ``Collaborative metric learning,'' in \emph{WWW}, 2017, pp. 193--201.

\bibitem{learningfeature}
C.~Xu and M.~Wu, ``Learning feature interactions with lorentzian factorization machine,'' in \emph{AAAI}, vol.~34, no.~04, 2020, pp. 6470--6477.

\bibitem{Learningcont}
M.~Nickel and D.~Kiela, ``Learning continuous hierarchies in the lorentz model of hyperbolic geometry,'' in \emph{ICML}.\hskip 1em plus 0.5em minus 0.4em\relax PMLR, 2018, pp. 3779--3788.

\bibitem{Tech_Rpt}
\BIBentryALTinterwordspacing
Lh-plugin technical report. [Online]. Available: \url{https://github.com/liujiaozhiren/LH-plugin}
\BIBentrySTDinterwordspacing

\bibitem{hnn}
O.~Ganea, G.~B{\'e}cigneul, and T.~Hofmann, ``Hyperbolic neural networks,'' \emph{NeurIPS}, vol.~31, 2018.

\bibitem{neutraj}
D.~Yao, G.~Cong, C.~Zhang, and J.~Bi, ``Computing trajectory similarity in linear time: {A} generic seed-guided neural metric learning approach,'' in \emph{ICDE}, 2019, pp. 1358--1369.

\bibitem{traj2s}
H.~Zhang, X.~Zhang, Q.~Jiang, B.~Zheng, Z.~Sun, W.~Sun, and C.~Wang, ``Trajectory similarity learning with auxiliary supervision and optimal matching,'' in \emph{IJCAI}, 2020, pp. 3209--3215.

\bibitem{st2vec}
Z.~Fang, Y.~Du, X.~Zhu, D.~Hu, L.~Chen, Y.~Gao, and C.~S. Jensen, ``Spatio-temporal trajectory similarity learning in road networks,'' in \emph{SIGKDD}, 2022, pp. 347--356.

\bibitem{chengdu}
{DiDi Research}, ``Didi gaia open dataset,'' \url{https://outreach.didichuxing.com/}, 2022.

\bibitem{porto}
{Kaggle}, ``Porto taxi trajectory dataset,'' \url{https://www.kaggle.com/competitions/pkdd-15-predict-taxi-service-trajectory-i/data}, 2015.

\bibitem{tdrive}
{Microsoft}, ``T-drive trajectory data sample,'' \url{https://www.microsoft.com/en-us/research/publication/t-drive-trajectory-data-sample/}, 2011.

\bibitem{osm}
\BIBentryALTinterwordspacing
O.~contributors, ``Openstreetmap,'' 2004, accessed: 2025-02-18. [Online]. Available: \url{https://www.openstreetmap.org}
\BIBentrySTDinterwordspacing

\bibitem{geolife}
\BIBentryALTinterwordspacing
M.~Research, ``Geolife trajectory dataset,'' 2008, accessed: 2025-02-18. [Online]. Available: \url{https://www.microsoft.com/en-us/download/details.aspx?id=52367}
\BIBentrySTDinterwordspacing

\bibitem{tedj}
\BIBentryALTinterwordspacing
D.~A. Tedjopurnomo, X.~Li, Z.~Bao, G.~Cong, F.~M. Choudhury, and A.~K. Qin, ``Similar trajectory search with spatio-temporal deep representation learning,'' \emph{{ACM} Trans. Intell. Syst. Technol.}, vol.~12, no.~6, pp. 77:1--77:26, 2021. [Online]. Available: \url{https://doi.org/10.1145/3466687}
\BIBentrySTDinterwordspacing

\bibitem{ndcg}
K.~J{\"a}rvelin and J.~Kek{\"a}l{\"a}inen, ``Cumulated gain-based evaluation of ir techniques,'' \emph{TOIS}, pp. 422--446, 2002.

\bibitem{dtw_para}
L.~Xiao, Y.~Zheng, W.~Tang, G.~Yao, and L.~Ruan, ``Parallelizing dynamic time warping algorithm using prefix computations on gpu,'' in \emph{2013 IEEE 10th International Conference on High Performance Computing and Communications \& 2013 IEEE International Conference on Embedded and Ubiquitous Computing}.\hskip 1em plus 0.5em minus 0.4em\relax IEEE, 2013, pp. 294--299.

\bibitem{erp}
L.~Chen and R.~T. Ng, ``On the marriage of lp-norms and edit distance,'' in \emph{VLDB}, 2004, pp. 792--803.

\bibitem{frechet}
T.~Eiter and H.~Mannila, ``Computing discrete fr{\'e}chet distance,'' 1994.

\bibitem{pioneer}
X.~Li, K.~Zhao, G.~Cong, C.~S. Jensen, and W.~Wei, ``Deep representation learning for trajectory similarity computation,'' in \emph{2018 IEEE 34th international conference on data engineering (ICDE)}.\hskip 1em plus 0.5em minus 0.4em\relax IEEE, 2018, pp. 617--628.

\bibitem{Hypergroup}
M.~Gromov, ``Hyperbolic groups,'' in \emph{Essays in group theory}.\hskip 1em plus 0.5em minus 0.4em\relax Springer, 1987, pp. 75--263.

\bibitem{treelike}
A.~B. Adcock, B.~D. Sullivan, and M.~W. Mahoney, ``Tree-like structure in large social and information networks,'' in \emph{ICDM}.\hskip 1em plus 0.5em minus 0.4em\relax IEEE, 2013, pp. 1--10.

\bibitem{hierar}
E.~Ravasz, A.-L. Barab{\'a}si, and Z.~Oltvai, ``Hierarchical organization of complex networks,'' in \emph{APS March Meeting}, vol. 2004, 2004, pp. V18--001.

\bibitem{hyper-vision}
A.~Ermolov, L.~Mirvakhabova, V.~Khrulkov, N.~Sebe, and I.~Oseledets, ``Hyperbolic vision transformers: Combining improvements in metric learning,'' in \emph{CVPR}, 2022, pp. 7409--7419.

\bibitem{hyper-learning}
J.~Yan, L.~Luo, C.~Deng, and H.~Huang, ``Unsupervised hyperbolic metric learning,'' in \emph{CVPR}, 2021, pp. 12\,465--12\,474.

\bibitem{hyper-img}
V.~Khrulkov, L.~Mirvakhabova, E.~Ustinova, I.~Oseledets, and V.~Lempitsky, ``Hyperbolic image embeddings,'' in \emph{CVPR}, 2020, pp. 6418--6428.

\end{thebibliography}

\end{document}